\documentclass[aps,pre,groupedaddress]{revtex4-1}
\usepackage{graphicx}
\usepackage{amsmath}
\usepackage{multirow}
\usepackage{color}
\usepackage{natbib}
\usepackage{hyperref}  

\begin{document}


\title{Magnetic induction and diffusion mechanisms in a liquid sodium spherical Couette experiment}


\author{Simon Cabanes}
\author{Nathana\"el Schaeffer}
\author{Henri-Claude Nataf}
\email[]{henri-claude.nataf@ujf-grenoble.fr}
\affiliation{Univ. Grenoble Alpes, ISTerre, F-38000 Grenoble, France}
\affiliation{CNRS, ISTerre, F-38000 Grenoble, France}
\affiliation{IRD, ISTerre, F-38000 Grenoble, France}


\date{\today}

\begin{abstract}
We present a reconstruction of the mean axisymmetric azimuthal and meridional flows in the DTS liquid sodium experiment.
The experimental device sets a spherical Couette flow enclosed between two concentric spherical shells where the inner sphere holds a strong dipolar magnet, which acts as a magnetic propeller when rotated.
Measurements of the mean velocity, mean induced magnetic field and mean electric potentials have been acquired inside and outside the fluid for an inner sphere rotation rate of $9$ Hz ($Rm \simeq 28$).
Using the induction equation to relate all measured quantities to the mean flow, we develop a nonlinear least square inversion procedure to reconstruct a fully coherent solution of the mean velocity field.
We also include in our inversion the response of the fluid layer to the non-axisymmetric time-dependent magnetic field that results from deviations of the imposed magnetic field from an axial dipole.
The mean azimuthal velocity field we obtain shows super-rotation in an inner region close to the inner sphere where the Lorentz force dominates, which contrasts with an outer geostrophic region governed by the Coriolis force, but where the magnetic torque remains the driver. 
The meridional circulation is strongly hindered by the presence of both the Lorentz and the Coriolis forces.
Nevertheless, it contributes to a significant part of the induced magnetic energy. 
Our approach sets the scene for evaluating the contribution of velocity and magnetic fluctuations to the mean magnetic field, a key question for dynamo mechanisms.
\end{abstract}


\maketitle


\section{Introduction\label{Introduction}}


The magnetic field of most planets and stars is generated by the dynamo effect, which converts kinetic energy into magnetic energy.
The apparition of a magnetic field is governed by the magnetic induction equation.
Solving this equation for a given velocity field is called the kinematic dynamo problem.
For a given velocity field of typical amplitude $U$ and characteristic length scale $L$, the magnetic field $B$ can grow only if the induction term is much larger than the diffusion term due to Ohmic dissipation.
This is measured by the magnetic Reynolds number $Rm =UL/\eta$, where $\eta$ is the magnetic diffusivity of the fluid.
Dynamo action requires a large enough magnetic Reynolds number.
However, it was soon recognized that this is not enough and that the velocity field must also satisfy particular topological properties.
For example, Cowling's theorem \cite{cowling33} states that a purely axisymmetric magnetic field will always decay.
Flows with a substantial helicity are usually more dynamo-prone, and the first experimental efforts concentrated on examples of such flows, for which the kinematic dynamo problem had been solved.
The team of Professor Gailitis built in Riga an experiment producing a swirling flow close to that of the Ponomarenko dynamo \cite{ponomarenko73}.
It used sodium (the best electric conductor known in the liquid state) as a working fluid.
In 2000, dynamo action was indeed observed \cite{gailitis01}, above a critical magnetic Reynolds number in agreement with the theoretical prediction.

In the mean time, the quest for a mechanism to produce a large-scale nearly axisymmetric magnetic field, such as observed on the Earth and other planets, led to the concept of $\alpha-\omega$ dynamos (see e.g. \citet{charbonneau10}).
In these models, a strong azimuthal flow around the rotation axis of the planet produces an $\omega$ effect, which shears any pre-existing poloidal magnetic field (such as a dipolar field), creating a strong azimuthal magnetic field.
In order to get around Cowling's theorem, non-axisymmetric flow structures are then required to convert some of this azimuthal field into a poloidal field and close the loop.
The small-scale non-axisymmetric flow structures that collectively produce a large-scale magnetic field can be viewed as one type of $\alpha$ effect.
This term has been introduced in the context of homogeneous isotropic hydrodynamic turbulence \cite{steenbeck66}.
Fluctuations also enhance the effective magnetic diffusivity, yielding a $\beta$ term that adds to $\eta$. 
This approach of small scale non-axisymmetric contributions to the mean magnetic field originates from the expansion of a mean electromotive force, $\epsilon = \langle \tilde u \times \tilde b\rangle$, in terms of the large-scale magnetic field $B$, such as $\epsilon = \alpha :  B  + \beta : \nabla \times  B $, \cite{krause80}. Components of the  $\alpha$ and $\beta$ tensors, which depend upon the
turbulent characteristics of the 
flow, have been evaluated from nonlinear numerical simulations  \cite{rudiger14} and in liquid metal experiments. In the latter case $\alpha$ and $\beta$, have been assumed to be scalars  \cite{frick10, rahbarnia12}. 

Experimental evidence for some sort of $\alpha$ effect was nicely demonstrated by the Karlsruhe dynamo \cite{stieglitz01}.
An array of 52 helicoidal flows of liquid sodium in pipes did produce a large-scale magnetic field above a critical value of the magnetic Reynolds number, as predicted by the theoretical analysis of G.O. Roberts \cite{roberts72}.
A large $\omega$ effect has been clearly observed in the liquid sodium experiment in New Mexico where a high toroidal field induction has been measured from a rotational shear in stable Couette flow \cite{colgate11}. Same has been observed in the DTS magnetized spherical Couette experiment \cite{brito11}, in which an imposed dipolar magnetic field is sheared by differential rotation of a liquid sodium layer (40 liters) between two concentric shells.
The complete $\alpha-\omega$ dynamo mechanism still awaits an experimental demonstration, but hints of such dynamo bursts have recently been observed in the 3m-facility of Dan Lathrop at the University of Maryland \cite{zimmerman14}.
This facility has the same geometry as the DTS experiment, but contains $12$ m$^3$ of liquid sodium.

Indeed, after the success of the Riga and Karlsruhe experimental dynamos, several teams have set up experiments to observe the dynamo mechanism in less constrained flows \cite{oconnell01, lathrop01, marie02b}.
A key element is that for the magnetic Reynolds number to be large enough for dynamo action, the kinetic Reynolds number $Re = UL/\nu$ has to be much larger, since the magnetic Prandtl number $Pm=\nu/\eta$ is of the order of $10^{-5}$ for liquid metals ($\nu$ is the kinematic viscosity).
None of these experiments has achieved self-excitation yet (except when ferromagnetic parts are present \cite{monchaux07, berhanu07}), and the main outcome of these studies is that turbulent fluctuations do indeed contribute to a large-scale magnetic field, but in a way that counters the action the mean large-scale flow alone would have \cite{spence06}.

It is thus of particular interest to investigate what are the actual contributions of turbulent small-scale fluctuations to magnetic induction and diffusion in various experimental conditions.
\citet{spence06} were the first to document a global negative $\alpha$ effect in the Madison facility.
Fluctuations were responsible for a $\sim 30\%$ reduction of the applied field for $Rm \simeq 130$.
\citet{frick10} investigated the effective magnetic diffusion in their torus experiment in Perm.
In this set-up, a torus filled with liquid sodium is spun to solid-body rotation and stopped abruptly. The authors applied a time-varying azimuthal magnetic field to the torus, and examined the magnetic response of the very turbulent flow that sets in just after the stop.
They report that turbulence increases the magnetic diffusivity by up to $\sim 30\%$ for $Rm \simeq 30$.
\citet{rahbarnia12} have recently performed a direct local measurement of the $\tilde{u} \times \tilde{b}$ term responsible for the $\alpha$ and $\beta$ effects (the $\tilde{ }$ denotes time-fluctuations).
They observe that the $\beta$ effect dominates and increases the effective diffusivity by $\sim 30\%$ for $Rm \simeq 160$.
Their results can be modeled as $\beta \simeq Rm_{rms}^*$, where $Rm_{rms}^* = u_{rms}\ell/\eta$ is a magnetic Reynolds number built on the length- and velocity- scales of the fluctuations.
\citet{nataf13} points out that the $\beta$ effect found by \citet{frick10} is almost ten times larger than that of \citet{rahbarnia12}, for a given  $Rm_{rms}^*$.
At this stage, we don't know if this is due to differences in the flow properties or to differences in the measurement methods.

In an effort to evaluate the contributions of turbulent fluctuations to the mean large-scale magnetic field in the DTS experiment, \citet{nataf13} inverted simultaneously mean velocity and magnetic data in order to evaluate by difference the part of the mean magnetic field that is not induced by the mean velocity field.
He also evaluated an upper bound for the contribution of the fluctuations to the mean field by mapping the amplitude of magnetic and velocity fluctuations.
He found that this contribution was too small to be resolved, but compatible with the results of \citet{rahbarnia12}.
Because his approach has the unique potential to provide maps of the small-scale contributions to the induction equation, we pursue our quest in this direction.

However, we improve the approach of \citet{nataf13} in a number of ways.
In particular, inspired by the method of \citet{frick10}, we analyse the response of the sodium layer to an imposed time-varying magnetic field.
This field simply results from small deviations of the spinning central magnet from axisymmetry.
We also extend the analysis to larger rotation rates (and hence larger magnetic Reynolds numbers).
This leads us to solve the full time-dependent kinematic induction equation.
Modifying and extending the code used by \citet{figueroa13}, we implement both the direct model (the induction equation) and the non-linear inverse problem.

The organization of this article is as follows:  we first recall the main characteristics of the DTS experiment.
We then present the non-axisymmetric magnetic signals, which constitute one of the main ingredients of our study.
We state the direct problem, with special care concerning the time-varying magnetic field, and describe the numerical computation of the kinematic non-linear problem.
Realistic conductivities and boundary conditions are considered.
We then perform a non-linear inversion of the experimental data at $Rm \simeq 28$ to recover the best axisymmetric azimuthal and meridional velocity fields.
The non-axisymmetric signals reveal a rich physics, which we unravel. 
We finally discuss the results and the perspectives for the evaluation of the contribution of fluctuations to the mean magnetic field.


\section{The DTS experiment\label{The DTS experiment}}

\subsection{The DTS device\label{The DTS device}}
The DTS experiment is a spherical Couette flow experiment, with forty liters of liquid sodium enclosed between two concentric spherical shells of respective radius $r_i^{\ast} = 74$ mm and $r_o^{\ast} = 210$ mm (the superscript $\ast$ refers to the dimensional quantities, and is dropped after adimensionalization by $r_o^{\ast}$ as given in Appendix \ref{Adimensionalization}).
The inner sphere consists of a $\simeq15$ mm-thick copper shell, which encloses a strong permanent magnet.
The stainless steel outer shell is $5$ mm thick.
The magnet produces a mainly dipolar field, pointing upward along the vertical rotation axis. Its intensity ranges from $B_i \simeq 180$ mT at the equator of the inner sphere, down to $B_o = 7.1$ mT at the equator of the outer sphere ($B_o$ is used for adimensionalization in Appendix \ref{Adimensionalization}).

We observe non-axisymmetric deviations of the imposed magnetic field, which reach a peak-to-peak amplitude of $16$ mT at the surface of the inner sphere, down to $0.1$ mT at the outer shell.
We will use these deviations as a novel tool to probe induction and diffusion in the sodium layer.
A reconstruction of the complete scalar magnetic potential of the magnet is presented in Appendix \ref{magnetic potential}.

The inner sphere can spin at rotation rates $f = \Omega/ 2 \pi$ up to 30 Hz, although we focus on measurements obtained for $f = 9$ Hz in this article.
The outer sphere can also rotate independently but we keep it at rest in the present case.
More details about the DTS device can be found in \citet{nataf06} and \citet{brito11}.

\subsection{Dimensionless numbers\label{Dimensionless numbers}}

This paper focuses on a kinematic model, which involves the induction equation alone.
Our system is hence governed by a single dimensionless number, the magnetic Reynolds number Rm.
Results presented in this article are for $Rm = 28$.

However the DTS experiment involves other dimensionless numbers related to the equation of motion.
We give in table \ref{tab:dimensionless numbers} the expressions and values of selected dimensionless numbers for two rotation rates of the inner sphere $f=9$ Hz and $30$ Hz.
The Elsasser number $\Lambda$ is classically used when a rotating flow interacts with a magnetic field.
It is a measure of the ratio of the Lorentz force over the Coriolis force.
In the DTS experiment, \citet{brito11} show that the transition between a Lorentz-dominated region near the inner sphere to a geostrophic outer region occurs at a cylindrical radius $s$ where the local Elsasser number (defined using $B(s)$ instead of $B_o$) is about 1.

The Hartmann number is larger than 180 everywhere in the sodium layer, indicating that magnetic forces dominate over viscous forces.
Since the magnetic Prandtl number Pm is very small, a large value of the Reynolds number Re is achieved for $Rm=28$.
We note that it is too large for present-day numerical simulations.
For this reason, a kinematic approach is of great interest as the equation of motion is solved by the experiment.
We also give the value of the Lundquist number Lu, which shows that Alfv\'en waves are severely damped by magnetic diffusion in our experiment.

 \begin{table}
 \caption{\label{tab:dimensionless numbers}}
 \begin{ruledtabular}
\begin{tabular}{ccccccccc}
{$f$} & {Rm} & $\Lambda$ & {Ha} & {Pm} & {Lu} &  {Re} \\
\colrule
$\Omega/ 2 \pi$ & $\Omega (r_o^\ast)^2 /\eta $ & $\sigma B_o^2 / \rho \Omega$ & $r_o^\ast B_o / \sqrt{\rho \mu \eta \nu}$ & $\nu / \eta$ & $r_o^\ast B_o / \sqrt{\rho \mu \eta^2}$  & $\Omega (r_o^\ast)^2 / \nu$ \\
\colrule
$9$ Hz & $28$ & $8.6 \cdot 10^{-3}$ & \multirow{2}{*}{$180$} & \multirow{2}{*}{$7.4 \cdot 10^{-6}$} & \multirow{2}{*}{$0.5$} & $3.8 \cdot 10^{6}$\\
$30$ Hz & $94$ & $2.6 \cdot 10^{-3}$ & &  & & $1.3 \cdot 10^{7}$\\
\end{tabular}
 \end{ruledtabular}
 \end{table}

\subsection{Axisymmetric data acquisition and processing \label{axisymmetric data acquisition and processing}}

At small magnetic Reynolds number and with an imposed magnetic field, one can obtain constraints on the fluid flow from the electrical currents and magnetic field it produces, as demonstrated by \citet{stefani00}. In the kinematic approach we follow, we use as many observational constraints as possible on the mean velocity field and on its electric and magnetic signatures.
We always assume that the mean flow is axisymmetric ($m=0$) and symmetric with respect to the equator.
We further assume that $f$ and $-f$ rotation rates yield the same meridional circulation and the opposite azimuthal flow.\\

\begin{figure}
\includegraphics[width=0.7\linewidth]{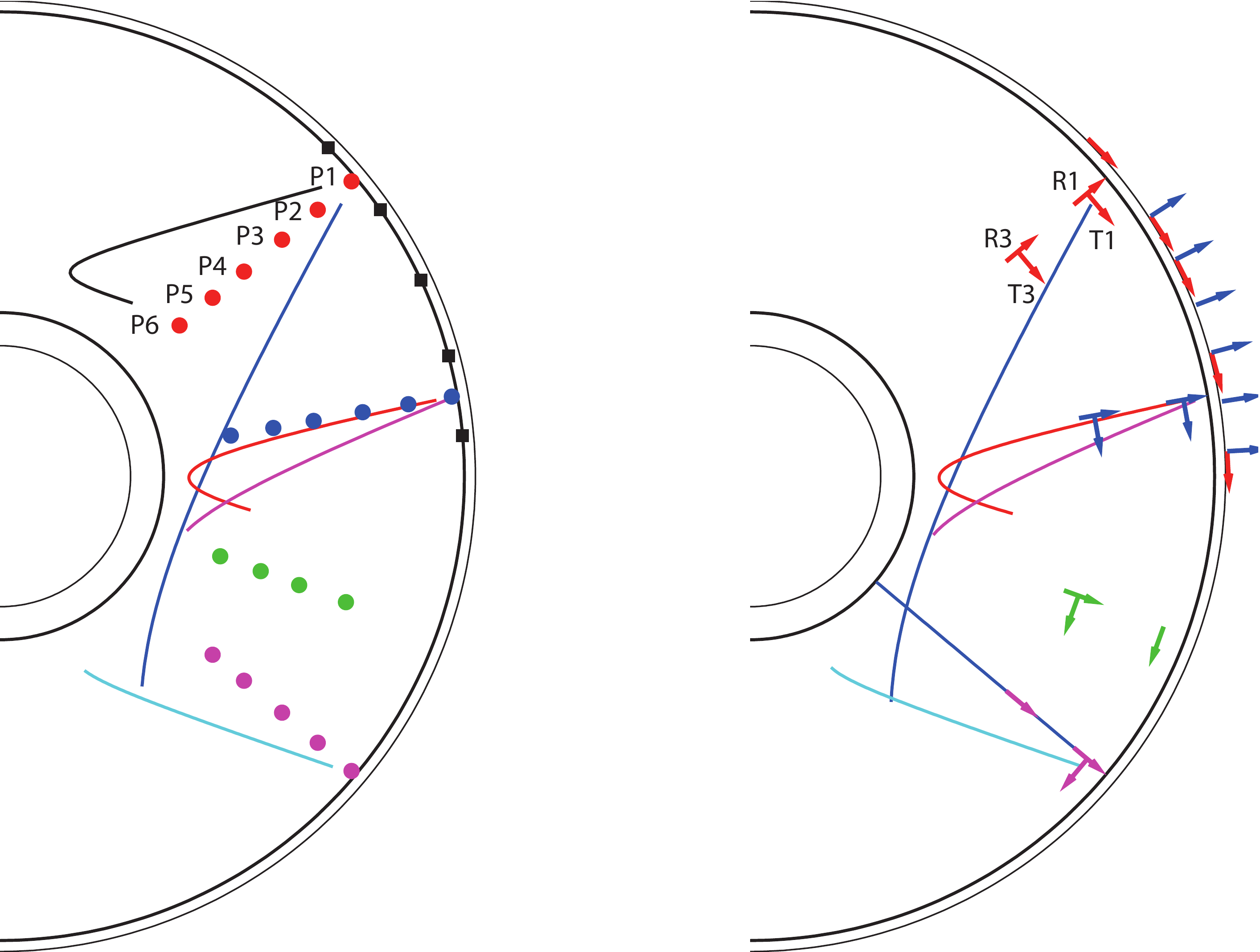}
\caption{(Color online) Meridional maps of data coverage. Black lines displaying four spherical shells, mark successively the copper, the fluid and the stainless steel shells.\textit{(left: azimuthal data)} The colored lines (crossing fluid shell) are the projections in the meridional plane of the ultrasound beams from which Doppler angular velocity profiles are obtained. Colored (gray) dots mark the position of the magnetic probes measuring the azimuthal component of the magnetic field ($B_{\varphi}$). Electric potentials are measured in the outer shell at positions displayed by the black squares. \textit{(right: meridional data)} The colored lines (crossing fluid shell) are the projections in the meridional plane of the ultrasound beams from which Doppler meridional velocity profiles are obtained. Colored (gray) arrows indicate the position of the magnetic probes measuring the radial and orthoradial components of the magnetic field inside and outside the fluid. The magnetic probes provide both the mean and the time-dependent signals used in this study.
\label{Azim_pt}}
\end{figure}

The $m=0$ data and data errors are acquired and processed as described by \citet{nataf13}.
The main difference is that we focus here on data for a rotation rate $f=\pm 9$ Hz, instead of $\pm 3$ Hz in \cite{nataf13}.
They include velocity profiles obtained by ultrasound Doppler velocimetry, $B_r$, $B_\theta$, and $B_\varphi$ measurements of the induced magnetic field inside the fluid shell, using Hall magnetometers arranged inside a sleeve that can be positioned at four different latitudes ($10^\circ$, $-20^\circ$, and $\pm40^\circ$), and electric potential differences at the surface of the outer sphere at four different latitudes ($10^\circ$, $20^\circ$, $30^\circ$, and $40^\circ$).
The resulting data coverage is displayed in Figure \ref{Azim_pt} in a meridional cross-section.
We also use the measured torque on the inner sphere, which can be related to the $l=2$ coefficient of the induced azimuthal magnetic field at the surface of the inner sphere, assuming the torque is essentially magnetic there \cite{nataf13}.
For the axisymmetric data, we make use of the $f \rightarrow -f$ symmetry.
We can thus obtain constraints on both the azimuthal flow and the meridional circulation from the sum and difference of Doppler velocity profiles shot along chords, which record a mixed projection.

\section{Non-axisymmetric signal\label{non-axisymmetric signal}}

Frequency power spectra of the magnetic field measurements are dominated by sharp peaks at the rotation frequency of the inner sphere and its multiple, especially in the sleeve close to the inner sphere.
These peaks are the signature of the small deviations of the imposed magnetic field from axisymmetry.
They record the response of the sodium layer to these deviations, which vary in time in the reference frame of the outer sphere since the non-axisymmetric components rotate with the inner sphere.
This opens the way to probing the effective magnetic diffusivity of the sodium layer, as pioneered by the group in Perm \citep{frick10}.

\subsection{Motifs and Fourier coefficients\label{sub:motifs}}

\begin{figure}[!h]
 \includegraphics[width=\linewidth]{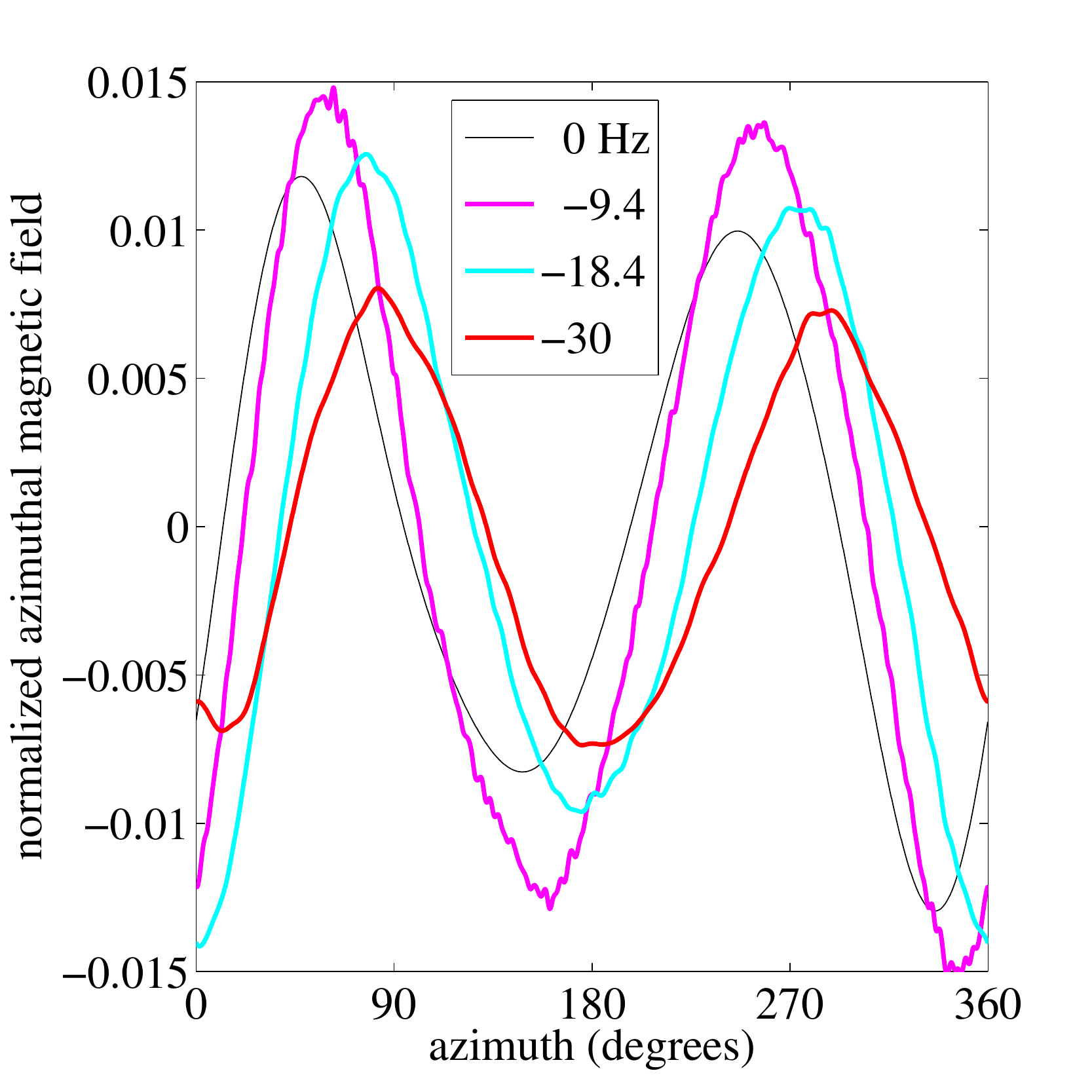}
 \caption{\label{motif}
(Color online) $2\pi$-longitudinal motifs of the non-axisymmetric magnetic field recorded by a Hall magnetometer measuring $B_\varphi$ for $r=0.79$ at a latitude of $-20^\circ$, for increasing rotation rates of the inner sphere $f=-9.4, -18.4$ and $-30$ Hz.
The magnetic field is normalized by $B_o$ as indicated in table \ref{tab:adimentionalization}.
The motif for $f=0$ is computed from the scalar magnetic potential of the magnet.}
\end{figure}

These entirely new data are obtained in two steps.
For each magnetometer, we first construct a $2 \pi$ longitudinal motif for each one-turn rotation of the inner sphere from the time series of the magnetic field recordings.
The records are chopped into $N$ one-turn-motifs, using a longitude marker (a small magnet glued on the pulley that entrains the inner sphere, which passes in front of a GMR magnetometer in the Lab reference frame).
$N$ is typically between $500$ for the lowest rotation rate up to $5000$ for the highest.
We assume that the angular velocity of the inner sphere is constant over one turn.
Figure \ref{motif} displays the mean motifs recorded by a given magnetometer at increasing rotation rates of the inner sphere.
One clearly sees the change in phase and amplitude that results from the response of the sodium layer.

The second step consists in performing a discrete Fourier transform of each of the $N$ motifs up to order $6$.
We then compute the mean and the root mean square $rms$ of the real and imaginary parts of each Fourier coefficient.
The mean and its standard deviation (estimated as $rms/\sqrt{N}$) are the data we use in the forthcoming inversion. Note that a systematic error has been added to the standard deviation to account for a possible misalignment of probes up to $5^{\circ}$.

\subsection{Non-axisymmetric data processing\label{non-axisymmetric data processing}}
Our data consists of the real and imaginary parts of the Fourier coefficients of the azimuthal motifs presented previously, with their associated standard deviations.
They are obtained from measurements of the induced magnetic field at various radii inside the sleeve, positioned at four different latitudes ($40^\circ$, $10^\circ$, $-20^\circ$, and $-40^\circ$), one at a time, as shown in Figure \ref{Azim_pt}.
Note that the data at $40^\circ$ and $-40^\circ$ are now treated separately since the non-axisymmetric part of the magnetic potential of the inner sphere's magnet is not symmetric with respect to the equator.
Similarly, we no longer use the $f \rightarrow -f$ symmetry, because the response of the fluid shell to the non-axisymmetric imposed magnetic field is not the same in these two cases, even though the velocity field keeps this symmetry.
The data and numerical modeling are therefore both retrieved for $f = -9$ Hz.

\section{Direct modelling of magnetic induction\label{direct model}}
\subsection{Numerical model}
In a fluid of magnetic diffusivity $\eta$, the magnetic field evolution is related to an imposed flow $\mathbf{U_0}$ by the induction equation:
\begin{equation}\label{F_induction}
\frac{\partial \mathbf{B}}{\partial t} = \mathbf{\nabla} \times ( \mathbf{U_0} \times \mathbf{B} ) + \eta \Delta \mathbf{B} 
\end{equation}
with
\begin{equation}\label{induction}
\nabla \cdot \mathbf{B} = 0,  \quad \quad \quad
\nabla \cdot \mathbf{U_0} = 0.
\end{equation}

We will solve this equation in the whole conducting domain, that is a spherical shell defined by $r_i-\delta_{Cu} < r < r_o+\delta_{ss}$, where $r_i$ and $r_o$ are the inner and outer radius of the shell filled with liquid sodium; $\delta_{Cu}$ is the depth of the copper inner shell in contact with the sodium and which encloses the permanent magnet; $\delta_{ss}$ is the depth of the stainless steel outer shell (see section \ref{The DTS experiment}).
We use a direct numerical simulation code of the induction equation, derived from the one used by \citet{figueroa13}. The fields are expanded in spherical harmonics \cite{schaeffer13}, and the equations are solved by finite differences in the radial direction.
The code has been improved to take into account sharp conductivity jumps (see appendix \ref{fd_jumps}) and time-dependent externally applied magnetic fields.
The magnetic field $\mathbf{B}(t)$ is evolved in time subject to diffusion, advection from a stationary and incompressible flow $\mathbf{U_0}$, and the rotating magnetic field boundary condition due to the magnet.
The real electric conductivities of the different shells are used.
A steady solution for the magnetic field is obtained after a few magnetic diffusion times, and we always wait 10 diffusion times before we compare to time averaged magnetic field and electric potential data collected on DTS.\\

We want to stress that the direct model solves for the full induction equation: there is no separation of applied and induced magnetic fields, and we do not assume any ordering between poloidal and toroidal fields.
Thus no large-scale induction terms are neglected and a complex non-linear dynamics can arise from interactions of the azimuthal and meridional components of both magnetic and velocity fields.
Taking into account all large-scale induction terms, we set the stage to quantify and map the part of the large-scale magnetic field that cannot be explained this way, and which could be produced by non-axisymmetric turbulent fluctuations.

\subsection{Field decomposition\label{Field decomposition}}
The velocity and magnetic fields are both divergence-free and can be decomposed into their poloidal and toroidal components.
\subsubsection{Velocity field\label{Velocity field}}
One can write for the velocity field, made dimensionless using time scale $\Omega^{-1}$ and length scale $r_o$ (see Appendix \ref{Adimensionalization}):

\begin{equation}\label{u1}
 \mathbf{U_0}= \nabla \times (U_T \, \mathbf{r})+ \nabla \times \nabla \times (U_P \, \mathbf{r})
\end{equation}
where $U_T$ and $U_P$ stand for toroidal and poloidal velocity scalar potential.
Since we only consider an axisymmetric mean velocity field, which does not depend upon the azimuthal component, and using the spherical coordinates $(r, \theta, \varphi)$ one simplifies \eqref{u1} such that
\begin{equation}\label{u2}
\mathbf{U_0}= 
\frac{1}{r} L_2(U_P) \, \vec{e}_r 
\: + \: \frac{1}{r} \partial_{\theta} \partial_r (r U_P) \, \vec{e}_{\theta}
\: - \: \partial_\theta U_T \, \vec{e}_{\varphi}
\end{equation}
 where $L_2$ is the angular laplacian. The two scalar fields $U_T$ and $U_P$ are expanded in spherical harmonics, leading to:
\begin{equation}\label{u3}
 U_{T}(r,\theta)= \sum_{odd \hspace {0.1cm}l}^{l_{max}}  u^T_l(r) Y^0_l (\theta)
\end{equation}
\begin{equation}\label{u4}
 U_{P}(r,\theta)= \sum_{even \hspace{0.1cm}l}^{l_{max}} u^P_l(r) Y^0_l (\theta)
\end{equation}
with $0< l \leq l_{max}$ the spherical harmonic degree.
We assume that the mean velocity field is symmetric with respect to the equator, so that only odd (respectively even) degree $l$ are considered for the toroidal (respectively poloidal) components.

To reduce the number of degrees of freedom of the inversion, the radial dependence of the flow is described using Tchebychev polynomials of the first kind $T_n$ of degree $n$.
Expressions \eqref{u3} and \eqref{u4} of toroidal and poloidal flow description become:
\begin{subequations}\label{uTcheb}
\begin{align}
 U_{T}(r,\theta) &= \sum_{odd \hspace {0.1cm}l}^{l_{max}} \sum_{n=0}^{n_{max}}  u^T_{ln} T_n(r) Y^0_l (\theta)\\
 U_{P}(r,\theta) &= \sum_{even \hspace{0.1cm}l}^{l_{max}} \sum_{n=0}^{n_{max}}  u^P_{ln} T_n(r) Y^0_l (\theta)
\end{align}
\end{subequations}
We choose $l_{max} = 8$, and $n_{max}=11$ so that $(n_{max}+1) \times l_{max} = 96$ free parameters describe the velocity field.

No-penetration of the fluid at the solid boundaries implies $U_P(r_i,\theta) =U_P(r_o,\theta) =0$ (which we will impose in the inversion).
In addition, realistic no-slip boundary conditions must be satisfied by the velocity field, leading to thin boundary layers, which cannot be resolved by our data and must be added to the previous radial description in order to run the forward model.
The velocity field has to be continuous on its outer boundary where the sphere is at rest:
$$U_T(r_o,\theta) =0$$
$$\partial_r U_P(r_o,\theta) = U_P(r_o,\theta) =0$$
and on its inner boundary where the azimuthal velocity matches that of the inner sphere $r_i \Omega \sin \theta$.
In order to satisfy these boundary conditions, we smooth the toroidal velocity profiles over a layer of thickness $\delta$ at the inner boundary as follows, for $l=1$:
\begin{equation}\label{ubc1}
u^{T*}_{1n}(r) = \left( u_{1n}^T(r_i) - u_{l=1}^T(r) \right) e^{-x_i} + u_{1n}^T(r)\\
\end{equation}
with $x_i = (r-r_i)/\delta$.
Similarly for $l>1$ at both boundaries, we write:
\begin{equation}\label{ubc}
u^{T*}_{ln}(r) =  u^{T}_{ln}(r) \left( 1 - e^{-x_o} - e^{-x_i} \right)
\end{equation}
with $x_o = (r_o-r)/\delta$.
We need both the poloidal component and its radial derivative to be smooth for all degrees $l$ while they both reach zero at the boundaries.
For this, we use:
\begin{subequations}\label{ubci}
\begin{align}
r < r_i + \delta & \quad \quad u^{P*}_{ln}(r) =  u^{P}_{ln}(r) \, x_i^3 (10 - 15 x_i + 6 x_i^2 ) \, ,\\
r > r_o - \delta & \quad \quad u^{P*}_{ln}(r) =  u^{P}_{ln}(r) \, x_o^3 (10 - 15 x_o + 6 x_o^2 ) \, ,\\
r_i+\delta \leq r \leq r_o-\delta  & \quad  \quad u^{P*}_{ln}(r) =  u^{P}_{ln}(r) \, .
\end{align}
\end{subequations}
Each $1$mm-thick boundary layer contains $20$ radial grid points.
Expressions \eqref{ubc1} to \eqref{ubci} are evaluated on a non equispaced radial grid of a total of $N_r = 300$ points and fed to the induction equation.
We have checked that the results of our inversions are almost independent of the size $\delta$ of the boundary layer.

\subsubsection{Magnetic field\label{Magnetic field}}
The same decomposition is used for the magnetic field:
\begin{equation}\label{eq:poltor}
 \mathbf{B}= \nabla \times (T \, \mathbf{r})+ \nabla \times \nabla \times (P \, \mathbf{r})
\end{equation} 
Where $T$ and $P$ stand for toroidal and poloidal magnetic scalar potentials. They are again projected on spherical harmonics:
\begin{subequations}\label{b22}
\begin{align}
 P= \sum_{m=0}^{M} \sum_{l=m}^{L} b^P_{lm}(r,t) \, Y^m_l (\theta, \varphi)\\
T= \sum_{m=0}^{M} \sum_{l=m}^{L}  b^T_{lm}(r,t) \, Y^m_l (\theta, \varphi)
\end{align}
\end{subequations}
Contrarily to the velocity field, we do not assume any symmetry for the magnetic field and retain all orders $m$, because we specifically want to include
measured deviations from the mostly dipolar field carried by DTS's permanent magnet.
The radial discretization including $r_i-\delta_{Cu} < r < r_o+\delta_{ss}$ is obtained on a non regular radial grid consisting of the same grid as the velocity field (300 points) extended in both solid shells with 50 more points.

We choose $L=11$ and $M=6$ here, which includes all of the experimental measurements. We checked that doubling $L$ did not change our results.

\subsection{Magnetic boundary condition\label{Magnetic boundary condition}}
In an insulator, where there is no electrical current ($\mathbf{j} = \nabla \times \mathbf{B} = 0$), the magnetic toroidal component vanishes ($T = 0$) and its poloidal component is a solution of $\Delta P = 0$.
Using the spherical harmonic decomposition \eqref{b22}b, the solutions are of the form:
\begin{equation}\label{bbc1}
b_{lm}^P = C_{lm} r^l + D_{lm} r^{-(l+1)}
\end{equation}
For the solution to remain finite, $C_{lm}$ and $D_{lm}$ are zero if there is no current or magnet for $r>r_o+\delta_{ss}$ or $r<r_i-\delta_{Cu}$, respectively.

Considering the outer interface between our conducting domain and the surrounding air ($\hat{r}_o=r_o+\delta_{ss}$), the magnetic field is continuous and so are  $P$ and $\partial_r P$. Indeed, on the outer interface, there is no externally imposed magnetic field, hence $C_{lm}$ vanishes, and by taking the derivative of expression \eqref{bbc1}, we find:
\begin{equation}\label{bbc2}
\left. \partial_r b_{lm}^P \right|_{\hat{r}_o} = - \frac{l+1}{\hat{r}_o} b_{lm}^P.
\end{equation} 
Similarly, at the inner boundary ($\hat{r}_i = r_i-\delta_{Cu}$), the poloidal magnetic field $P$ and its derivative $\partial_r P$ are still continuous.
However, one must now take into account the magnetic field produced by the magnet, so that the resulting magnetic field at the interface is both fed by an inner and an outer magnetic source.
Hence, both $C_{lm}$ and $D_{lm}$ are non zero in expression \eqref{bbc1}.
Since we know that the magnetic field imposed by our magnet is defined by ${b_0}_{lm}^P(r, t) = D_{lm}(t) \, r^{-(l+1)}$, we can rewrite the radial derivative of equation \eqref{bbc1} at the inner boundary as:
\begin{equation}\label{bbc3}
\left. \partial_r b_{lm}^P \right|_{\hat{r}_i} = \frac{l}{\hat{r}_i}b_{lm}^P - \frac{2l+1}{\hat{r}_i} {b_0}_{lm}^P(t)
\end{equation}
where the time dependent boundary condition is related to the inner sphere angular velocity $\Omega$ by
$${b_0}_{lm}^P(t) = {b_0}_{lm}^P(0) \, e^{im \Omega t}.$$
The imposed magnetic field at the inner boundary is thus fully described by the complex spherical harmonic coefficients ${b_0}_{lm}^P(0)$ computed from the measured scalar magnetic potential of the magnet presented in Appendix \ref{magnetic potential}.

\section{Inversion\label{Inversion problem formulation}}
\subsection{Non linear inversion formalism\label{general nonlinear ls}}
The data acquisition procedure, detailed in sections \ref{axisymmetric data acquisition and processing} and \ref{non-axisymmetric data processing}, provides a heterogeneous data set, which comprises both axisymmetric and non-axisymmetric data.
Let's call $d$ the vector collecting these $N$ data points measured at the positions displayed in Figure \ref{Azim_pt}.
For a given velocity field, the kinematic approach described in section \ref{direct model} solves the induction equation  \eqref{F_induction}, and provides predictions of the data.
We call $m$ the model vector, which collects the $96$ coefficients $u_{ln}^T$ and $u_{ln}^P$ of the velocity field according to the decomposition \eqref{uTcheb}. 
In order to carry out the inversion and find the best velocity model, one wants to minimize a cost function consisting of the difference between data vector $d$ and  prediction vector $g(m)$ weighted with the error, plus the deviation from an \textit{a priori} model (with weight detailed in the covariance matrix below).

We note that even if the direct model $g(m)$ is linear for a given velocity field, such is not necessarily the case for the inverse problem.
Indeed, at moderate $Rm$ value, a non linearity arises from the cross-product of the velocity solution and an induced magnetic field, which also depends on the velocity field.
Under such conditions, the inverse problem has to be considered non linear and the best fitting model is obtained using the classical generalized nonlinear least square inverse method \cite{tarantola82}:
\begin{widetext}
\begin{equation}\label{least square}
m_k = \left( G_{k-1}^T C_{dd}^{-1} G_{k-1} + C_{pp}^{-1} \right)^{-1} \bigg \{ G_{k-1}^T C_{dd} \big \{ d-g(m_{k-1}) \big \} - C_{pp}^{-1} (m_{k-1} - m_0) \bigg \}
\end{equation}
\end{widetext}
where $C_{dd}$ is the covariance matrix of the data and $C_{pp}$ the \textit{a priori} covariance matrix of the model parameters, which we describe below.
The matrix $G$ collects the Fr\'echet derivatives of $g(m)$ and is detailed below.
We index by $k$ the iterative process which leads from the initial model $k=1$ to the final model, for which the cost function is minimal.
At each step $k$, the normalized misfit $\chi_k$ is defined as:
\begin{equation}\label{misfit}
\chi_k = \sqrt{\frac{1}{N} \sum_n^N \left(\frac{d-g(m_k)}{\sigma_d} \right)^2}
\end{equation}
where $\sigma_d$ account for the experimental error of each data $d$.\\


At each step $k$, the element of matrix $G_k$ for data point $i$ and parameter $j$ is given by,
\begin{equation}\label{G}
G_k(i,j) =  \frac{\partial g_i(m_k)}{\partial m_j} \simeq \frac{g_i(m_k(j) + \delta m_j) - g_i(m_k(j))}{\delta m_j},
\end{equation}
where $\delta m_j$ is a small step in parameters space chosen as $\delta m_j = 5\% \times m_k(j)$. \\

The covariance matrix of the data $C_{dd}$ is taken diagonal.
The diagonal terms are the square of the standard deviations presented in the data processing sections \ref{axisymmetric data acquisition and processing} and \ref{non-axisymmetric signal}.
We use the \textit{a priori} covariance matrix of the model in order to smooth the model in the radial and latitudinal directions.
This smoothness is controlled by the truncation degrees $l_{max}$ and $n_{max}$ of the \textit{Legendre} and  \textit{Tchebychev} expansions, but also by the covariance matrix in which the highest spectral degrees are damped, using:
\begin{equation}\label{cpp}
C_{pp} = \frac{c}{l^{a} + n^{a}} I ,
\end{equation}
where $I$ is the identity matrix.
We choose $a=4$ and $c=0.05$ to avoid undesired radial and latitudinal oscillations of the solution.

\subsection{Inverted model for $f = -9$ Hz\label{profiles and predictions}}

\subsubsection{Inversion convergence}
We present the results of a simultaneous inversion of the poloidal and toroidal mean velocity fields for an inner sphere rotation rate $f = -9$ Hz.
The data we invert for includes ultrasound Doppler velocity profiles, the induced magnetic field in the sleeve with both its mean axisymmetric ($m=0$) and time-dependent non-axisymmetric ($1\leq m \leq 5$) components,  the electric potential differences at the surface and the torque on the inner sphere.
Details of the data coverage are presented in section \ref{The DTS experiment}.

We perform the non-linear iterative process described in section \ref{general nonlinear ls} and we summarize in table \ref{tab:misfits} the global normalized misfit $\chi_k$ at each step $k$, as well as the detailed misfit for each data type.
We note that the minimum is reached after six iterations and that the misfit decreases strongly at the first iteration.
Some non linearity is present since the misfit of some data go through successive minimum and maximum.
Magnetic torque $\Gamma_M$ is an example of such a behavior as it successively reaches two local minima.

 \begin{table}
 \caption{Normalized misfits obtained from expression \eqref{misfit} and computed at each iteration $k$ of our inversion process.
We detail the misfit of the axisymmetric and non-axisymmetric data described in sections \ref{axisymmetric data acquisition and processing} and \ref{non-axisymmetric data processing} as well as ultrasound Doppler velocity profiles. $\chi_k$ represents the global normalized misfit.
 \label{tab:misfits}}
\begin{ruledtabular}
\begin{tabular}{| c | c | cc|cc|c|c|}
			 \multirow{2}{*}{$\; k\; $}  & {Non-axisymmetric} & \multicolumn{2}{c|}{Doppler profiles}  & \multicolumn{2}{c|}{m=0 magnetic field} &  \multirow{2}{*}{$\Gamma_M$} &  \multirow{2}{*}{$\chi_k$}\\
						     & {magnetic field}		        & 	{azim} & {merid}	       		 & {azim} & {merid} &   & \\
\colrule
\colrule
							$1$&$4.08$ & $10.62$ & $1.53$ & $3.87$ & $0.99$  & $24.09$ & $8.74$\\
							$2$&$3.18$ & $\; 1.47$ & $0.72$ & $3.63$ & $1.54$  & $\; 0.63$ & $1.67$\\
							$3$&$3.18$ & $\; 1.47$ & $0.65$ & $3.74$ & $1.26$  & $\; 1.19$ & $1.66$\\
							$4$&$3.15$ & $\; 1.48$ & $0.67$ & $3.67$ & $1.15$  & $\; 0.26$ & $1.65$\\
							$5$&$3.16$ & $\; 1.48$ & $0.70$ & $3.65$ & $2.35$  & $\; 0.81$ & $1.68$\\
							$6$&$3.17$ & $\; 1.47$ & $0.66$ & $3.64$ & $1.06$  & $\; 0.52$ & $1.65$\\

\end{tabular}
 \end{ruledtabular}
 \end{table}

Note that we used as an initial solution a velocity field obtained from a previous inversion at $f = -3$ Hz, from \citet{nataf13}, rescaled to the present rotation rate.
We thus converge more rapidly to the best fit.
Moreover, it should be realized that the phase of the non-axisymmetric magnetic data is defined \textit{modulo} $2 \pi$.
Thus, non linearity can also emerge from solutions with a large azimuthal shear, which can stretch magnetic heterogeneities in such a way that the phase gets an extra $\pm 2\pi$.
The use of a good initial guess is our way to handle that sort of non linearity.

\subsubsection{Radial profiles\label{Radial profiles fields description}}
\begin{figure}
 \includegraphics[width=\linewidth]{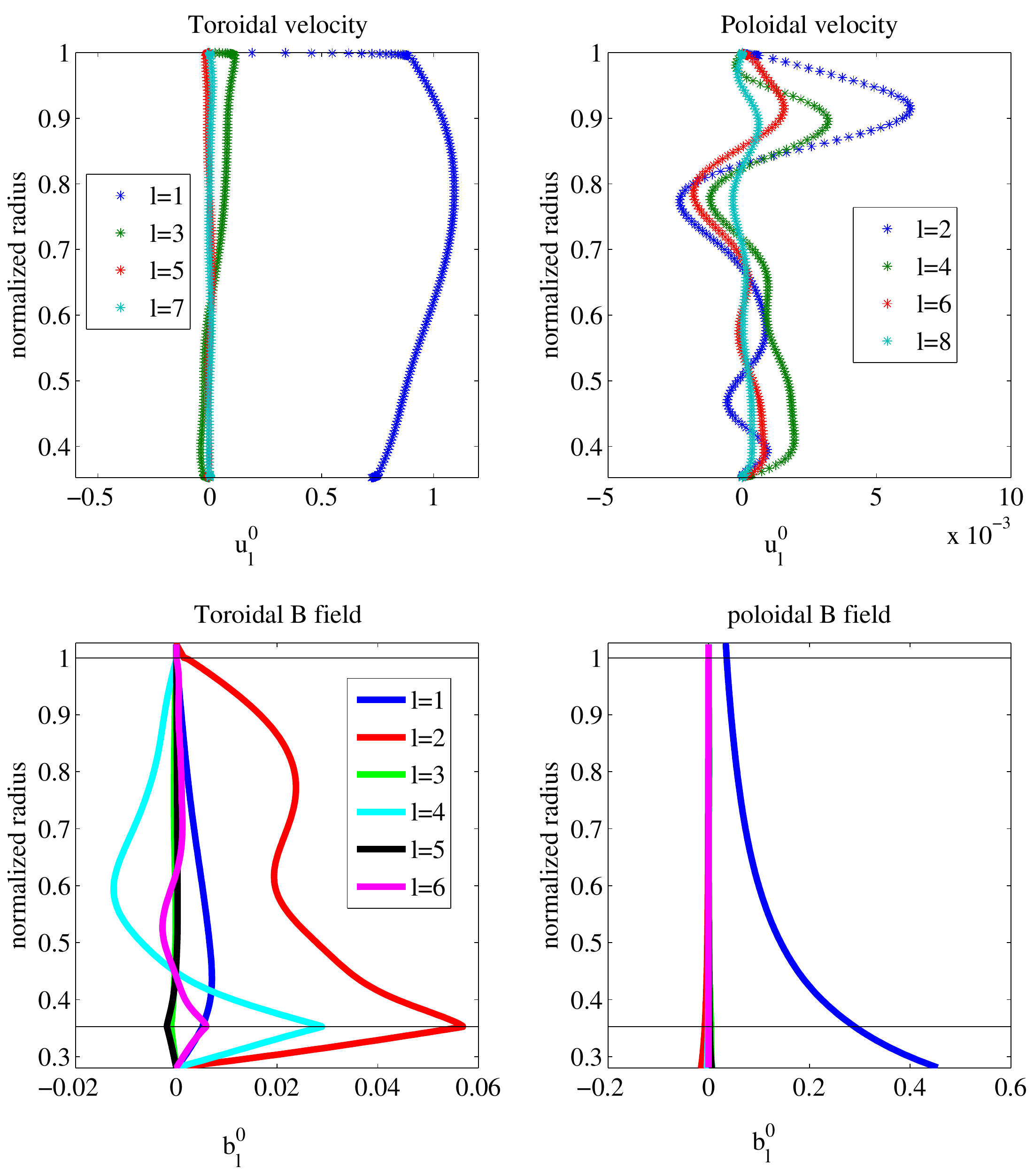}
   \caption{(Color online) Radial profiles of velocity and magnetic modes for all harmonic degrees. Upper plots from left to right: toroidal $u_l^T(r)$ and poloidal $u_l^P(r)$ velocity modes. Bottom plots from left to right: axisymmetric toroidal $b^T_{l,m=0}(r)$ and poloidal $b^P_{l,m=0}(r)$ magnetic field modes. All fields are normalized as given in Appendix \ref{Adimensionalization}. The radius axis of the magnetic plots extends from $r_i-\delta_{Cu}$ to $r_o+\delta_{ss}$ and horizontal lines indicate the fluid/solid interfaces at $r_i$ and $r_o$.
\label{Tp_c}}
\end{figure}

We show in Figure \ref{Tp_c} the radial profiles of the toroidal and poloidal velocity modes, $u^T_l(r)$ and $u^P_l(r)$ (top profiles), and of the axisymmetric magnetic modes, $b^T_{lm=0}(r) $ and $b^P_{lm=0}(r)$ (bottom profiles), obtained by our inversion.\\

\begin{figure}
 \includegraphics[width=\linewidth]{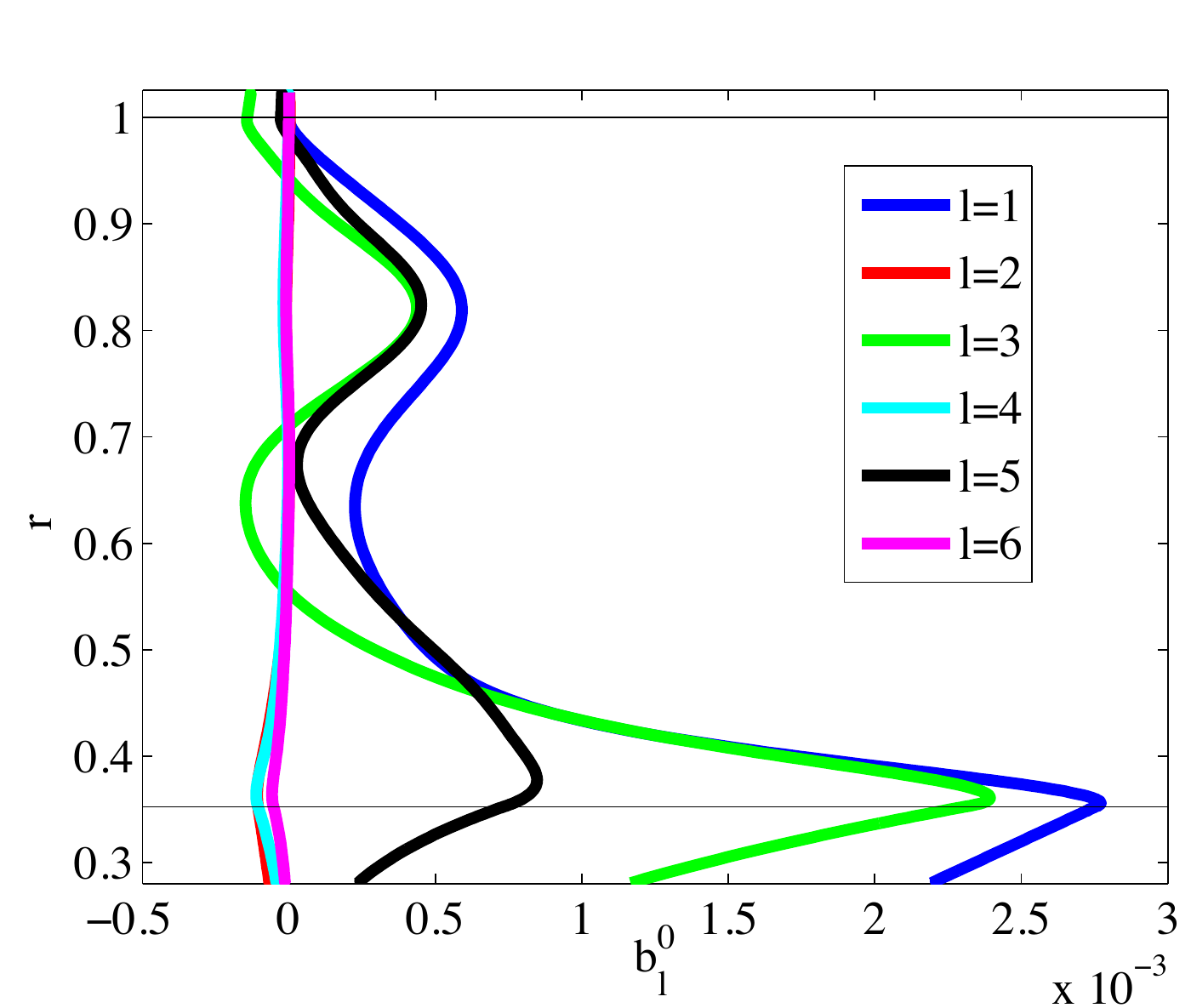}
    \caption{(Color online) Radial profiles of the induced axisymmetric poloidal magnetic modes for all harmonic degrees.
Each latitudinal degree from $l=1$ to $l=6$ is represented using color lines (grayscale) depicted in the legend.
The field is made dimensionless as given in Appendix \ref{Adimensionalization}. The radius axis extends from  $r_i-\delta_{Cu}$ to $r_o+\delta_{ss}$. and horizontal lines indicate the fluid/solid interfaces at $r_i$ and $r_o$.  \label{Dinduc_c}}
\end{figure}

We note that the toroidal (or azimuthal) degree $l=1$ largely dominates the velocity field.
The poloidal (or meridional) velocity, which is two orders of magnitude lower than the toroidal velocity, is distributed on latitudinal degrees $l=2,4,6$.
Note that the poloidal velocity profiles are less smooth than the toroidal ones.
Concerning the magnetic field, Figure \ref{Tp_c} shows an important toroidal induction close to the inner sphere, which is dominated by modes $l=2$ and $4$.
This magnetic induction is the signature of the strong azimuthal shear of the magnetic field lines known as the $\omega$-effect.
The induced toroidal magnetic field rapidly drops to zero as it diffuses in the solid copper shell where no induction occurs.
The bottom right plot of Figure \ref{Tp_c} shows that the poloidal magnetic field is largely dominated by the axisymmetric dipole ($l=1, m=0 $) imposed by the permanent magnet.
We access the axisymmetric induction of the poloidal magnetic field on Figure \ref{Dinduc_c} by subtracting the imposed field.
We observe that poloidal induction is especially strong near the inner sphere for odd degrees, including the dipole ($l=1$) and degrees $l=3$ and $5$ while the even degrees are negligible (close to zero value).
In the absence of non-linear poloidal-to-toroidal coupling in the induction term, all odd degrees would vanish for the toroidal magnetic field, and all even degrees for the poloidal magnetic field.
It is finally interesting to note that the induced axisymmetric dipole vanishes at the outer boundary (Figure \ref{Dinduc_c}), in agreement with \citet{spence06} who demonstrated that no dipole moment can be induced in a simply connected axisymmetric system.

\subsubsection{Meridional maps\label{fields Maps}}

\begin{figure}
\includegraphics[width=0.5\linewidth]{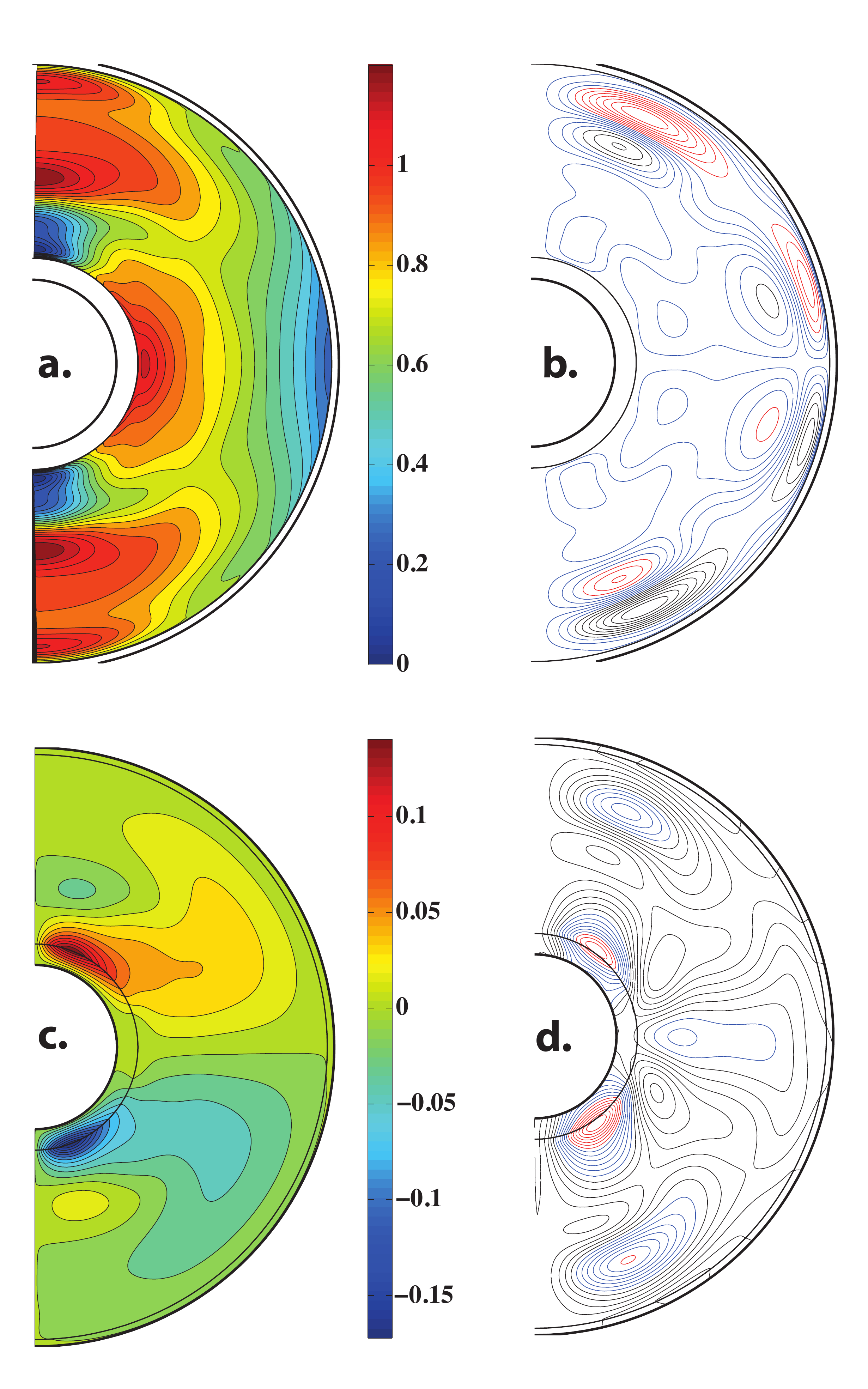}
     \caption{(Color online) Maps of the velocity and magnetic fields in a meridional $(r, \theta)$ section. Black spherical lines mark successively the copper, the fluid and the stainless steel shells.
\textbf{a.} Contour map of the angular velocity.
\textbf{b.} Stream lines of the meridional flow.
The flow is centrifugal along the equator, and values of the stream function range from  $-4 \cdot 10^{-4}$ to $4 \cdot 10^{-4}$.
\textbf{c.} contour map of the azimuthal magnetic field ($B_{\varphi}$) at longitude $\varphi=0$.
\textbf{d.} Field lines of the induced meridional magnetic field at $\varphi=0$.
\label{fields}
}
\end{figure}

Figure \ref{fields} shows maps of the velocity and magnetic fields in a meridional $(r, \theta)$ section.
Figure \ref{fields}a is a contour map of the fluid angular velocity $\omega$, while figure \ref{fields}b displays the streamlines of the meridional circulation (recall that both are assumed axisymmetric).
The lower maps are the magnetic field maps at longitude $\varphi = 0$.
Figure \ref{fields}c is a contour map of $B_{\varphi}$, and figure \ref{fields}d shows the field lines of the induced poloidal magnetic field.
All quantities are dimensionless, as given in Appendix \ref{Adimensionalization}.

Most of the features depicted by the angular velocity map (figure \ref{fields}a) have already been discussed by \citet{brito11}, and by  \citet{nataf13} for a lower rotation rate $f = \pm3$ Hz of the inner sphere.
We confirm that the fluid is entrained very efficiently by the magnetic inner sphere.
Super-rotation ($\omega > 1$) is observed in the equatorial region close to the inner sphere, where the flow obey Ferraro's law of isorotation \cite{ferraro37}.
Further away from the inner sphere, the magnetic stress is reduced and the Coriolis force dominates.
The flow is then geostrophic and the angular velocity shows little variation along the rotation axis.
We observe that super-rotation only reaches $10\%$ for $f=-9$ Hz, as compared to more than $20\%$ for $f=\pm3$ Hz \cite{brito11, nataf13}.
Together with the shift of that zone closer to the inner sphere, this indicates that the geostrophic region extends further inwards at larger rotation rates, in agreement with the observations of \citet{brito11}.
We remark that the polar region is poorly constrained by the data for technical reasons reported in section \ref{The DTS experiment}, and hence the flow near the axis should not be trusted.

The meridional circulation we obtain at $f=-9$ Hz (figure \ref{fields}b) differs more strongly from that derived by \citet{nataf13} at $f=\pm3$ Hz.
We still get a cell that drives fluid toward the equator along the inner sphere, but strong counter-rotating cells are active both beneath the outer surface and at high latitudes on the inner sphere.
Overall, our (normalized) meridional velocities are one order of magnitude larger than the ones obtained by \citet{nataf13} at lower rotation rate ($f = 3$ Hz).

The associated induced magnetic field is dominated by azimuthal induction (figure \ref{fields}c), which mainly occurs in two distinct regions: one located at the inner sphere surface and the other one in the geostrophic shear region.
Note that the induced magnetic field maps are not perfectly symmetric with respect to the equator, as our model includes the measured heterogeneities of the imposed magnetic field.

\subsubsection{Energies}

In order to better quantify the evolution of the flow and induced magnetic field with the Reynolds number, we compare in table \ref{tab:energy} the various energies computed from our model at $Rm=28$ with those of \citet{nataf13} for $Rm=9.4$.
Energies are obtained by integration of the magnetic and velocity solutions over the fluid shell. 
All quantities are scaled using $\rho \eta^2 r_o^{\ast} Rm^2$.
We note that the magnetic energy of the system is largely dominated by the imposed magnetic dipole $E_M^{dipole}$, which is in fact independent of the spin rate of the inner sphere. \\

We observe that the azimuthal kinetic energy decreases from $E_K^{Tor} = 0.45$ at $Rm=9.4$ to $E_K^{Tor} = 0.33$ at $Rm=28$.
It corresponds to the reduction of the super-rotation zone described in the previous section \ref{fields Maps}.
The geostrophic shear thus extends further into the strong magnetic field region and hence increases the azimuthal magnetic energy to $E_M^{Tor} = 3.3 \cdot 10^{-4}$ instead of $E_M^{Tor} = 1.7 \cdot 10^{-4}$ at lower Rm.

Recent numerical simulations of the non-magnetized spherical Couette flow in our geometry \cite{wicht14} yield a poloidal to toroidal kinetic energy ratio of about $18\%$.
Our model only yields $5\%$, while it barely reached $\approx 0.1 \%$ in Nataf's model \cite{nataf13}.
However, this relatively sluggish meridional flow induces a strong meridional magnetic field, yielding a poloidal to toroidal magnetic energy ratio of $0.35$.

 \begin{table}
 \caption{Energies of the imposed dipolar field, the induced toroidal and poloidal magnetic fields, and of the toroidal and poloidal flows. We report energies for two models at different Reynolds numbers. Model \textbf{(a)} corresponds to the kinematic model of the present paper at $Rm=28$ and \textbf{(b)} is a model at $Rm=9.4$ from \citet{nataf13}. All energies are dimensionless, as given in Appendix \ref{Adimensionalization}. We also report the poloidal to toroidal radio of the kinetic and magnetic energies. \label{tab:energy}}
 \begin{ruledtabular}
\begin{tabular}{c|ccccc|ccc}
			\textrm{Model} & {$E_M^{dipole}$}  & {$E_M^{Tor}$}  & {$E_M^{Pol}$}  & {$E_K^{Tor}$}  & {$E_K^{Pol}$} & {$E_M^{Pol}/E_M^{Tor}$} & {$E_K^{Pol}/E_K^{Tor}$} \\
\hline
								{\textbf{(a)} Our model} & $23/Rm^2$ & $3.3 \cdot 10^{-4}$ & $1.1 \cdot 10^{-4}$ & $0.33$ & $0.017$ &  $0.35$ & $0.051$\\
								{\textbf{(b)} \citet{nataf13}} & $23/Rm^2$ & $1.7 \cdot 10^{-4}$ & $2.0 \cdot 10^{-5}$ & $0.45$ & $4 \cdot 10^{-4}$  & $0.11$ & $0.001$\\

\end{tabular}
 \end{ruledtabular}
 \end{table}
 
\subsection{Axisymmetric data predictions\label{Data predictions}}
We now confront the predictions of the full induction equation model, involving cooperative effects of the meridional and azimuthal flows, with the experimental measurements used for inversion.


\begin{figure}
\includegraphics[width=0.7\linewidth]{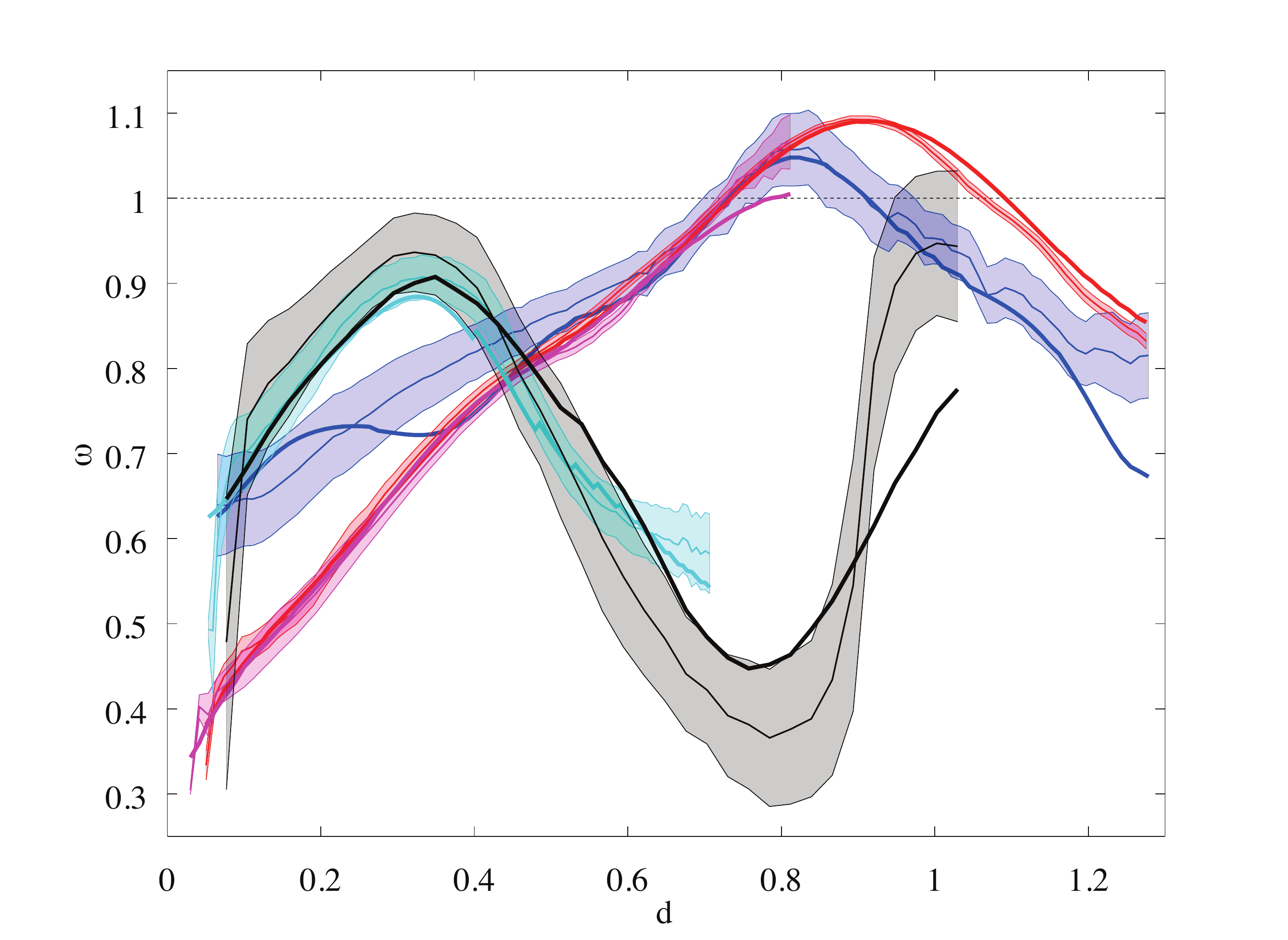}
      \caption{(Color online) Fits of the angular velocity predictions to ultrasound velocimetry doppler profiles. Measured data are thin lines presented with their error bars. Thick lines are predictions. All profiles are function of the distance $d$ along the ray of the ultrasound beam. Colors (grayscale) refer to ultrasound beam path at different latitudes as displayed on data coverage map Figure \ref{Azim_pt}.  All data are dimensionless as given in Appendix \ref{Adimensionalization}. \label{dop_}}
\end{figure}
\begin{figure}
 \includegraphics[width=\linewidth]{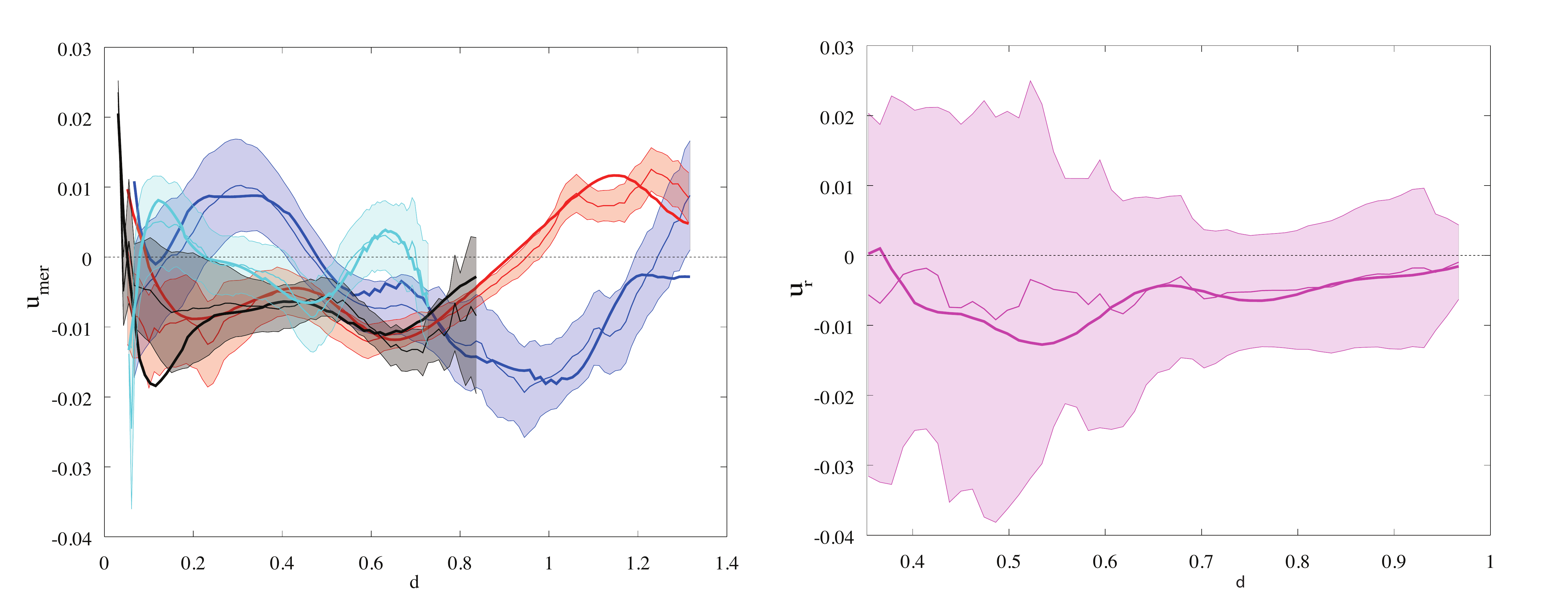}
      \caption{(Color online) Fits of velocity predictions to ultrasound velocimetry doppler profiles. These plots compare the measured data with their error bars to our kinematic model predictions of the meridional (left) and radial velocity (right). Both profiles are function of the distance $d$ along the ray of the ultrasound beam. All data are dimensionless as given in Appendix \ref{Adimensionalization}. \label{dop_merid}}
\end{figure}
We compare our model's predictions to the ultrasound Doppler profiles of the angular velocity on Figure \ref{dop_} and of the meridional and radial velocity on Figure \ref{dop_merid}. The four ultrasound Doppler profiles describing the angular velocity are rather well explained by the model. Except the blue (lightest gray) profile near the outer sphere and around $d=0.4$ and the black profile close to the inner sphere, all other velocity predictions are within the error bars.
We consequently obtain a  normalized misfit which reaches a value slightly greater than one ($\chi = 1.47$).
The meridional and radial velocity predictions fall perfectly in error bars giving a normalized misfit of $0.65$. We note that despite the large error bars carried by the radial profile, model predictions fall very close to the mean value.

\begin{figure}[!h]
 \includegraphics[width=\linewidth]{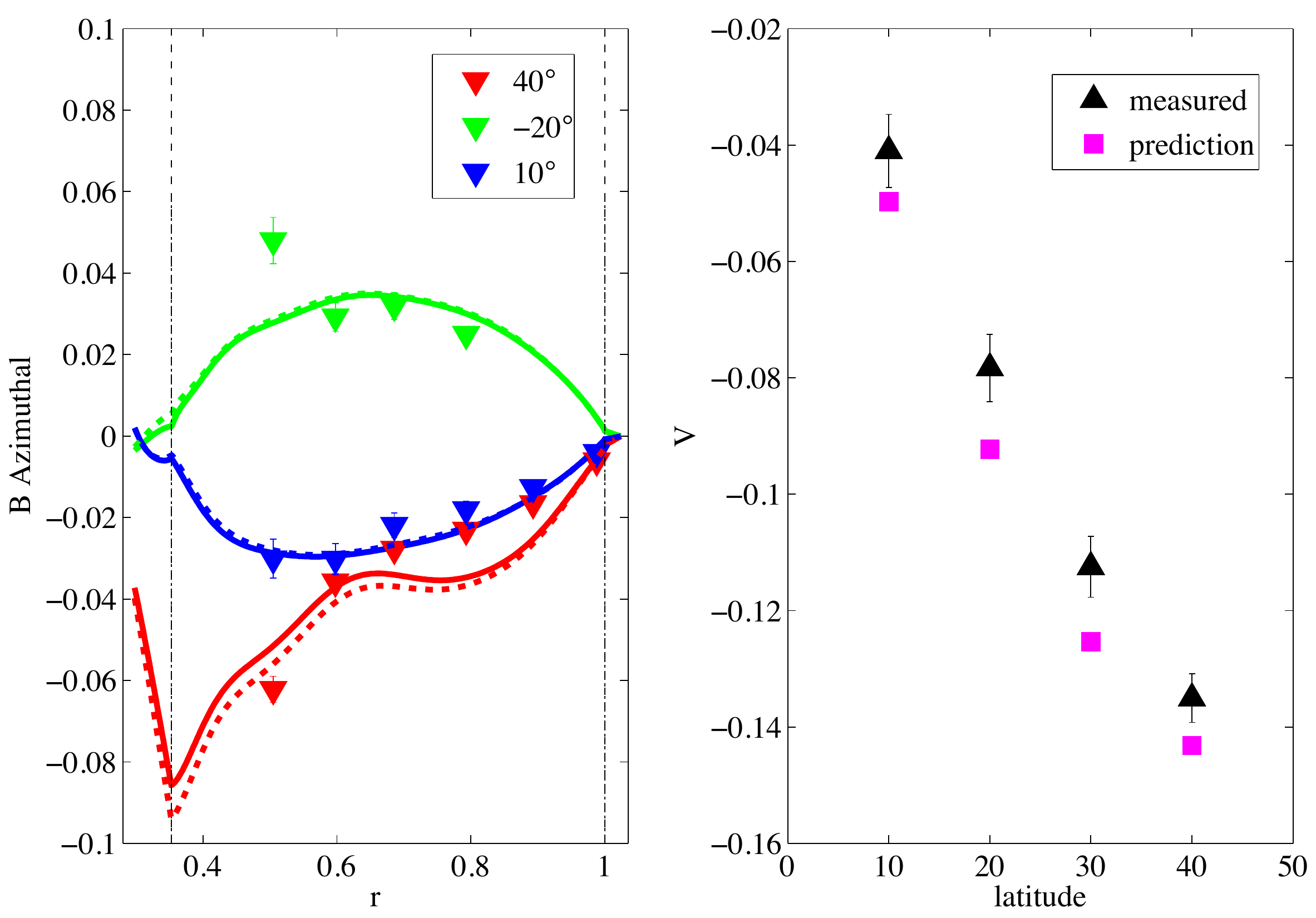}
       \caption{(Color online) Azimuthal predictions and measurements: {\it(left: azimuthal magnetic field)} azimuthal  magnetic field predictions inside the fluid with the measured data and their error bars.
The triangles with error bars are our measurements at three latitudes: $40^{\circ}$ (bottom red markers), $10^{\circ}$ (middle blue markers) and $-20^{\circ}$ (upper green markers).
The solid lines are our model predictions at the same latitudes.
The velocity model contains both the azimuthal and the meridional flows.
The dashed lines are the predictions for the same flow without meridional circulation.
The radius axis extends from $r_i-\delta_{cu}$ and $r_o+\delta_{ss}$ and vertical dash-dot lines indicate the fluid/solid interfaces at $r_i$ and $r_o$.
{\it(right: surface electric potential)} the black triangles with error bars are the differences in electric potential between electrodes $10$ degrees apart that we measure at the surface of the outer shell.
The red squares are our model predictions.
All fields are dimensionless as given in Appendix \ref{Adimensionalization}.
\label{fits_b}}
\end{figure}

Let's now see on Figure \ref{fits_b} how well the kinematic model predicts the induced azimuthal  ($m=0$) magnetic field inside the sodium shell and the electric potential differences measured at the outer sphere surface. 
Figure \ref{fits_b} shows reasonable agreements between azimuthal magnetic field (which is mainly induced by the azimuthal velocity) predictions and observations, leading to the conclusion that induction from the mean flow do an important part of the job.
The high latitudes measurements are not well explained (red (bottom) curve of Fig. \ref{fits_b}) especially between $r=0.6$ and $r=0.9$. A plausible explanation could be the contribution from turbulent fluctuations to our mean field.
In addition, the innermost measurements (for $r<0.6$) often show significant discrepancy. The misfit for the azimuthal magnetic field is $3.64$.
Dashed lines on Figure \ref{fits_b} display the induced azimuthal magnetic field by the azimuthal flow only.
It shows that the induced field increases when no meridional flow is involved. We note that this meridional advection effect increases with the latitude and shows a deviation of the induced azimuthal magnetic field of about $10 \%$ at  latitude $40^{\circ}$.

Electric potential at the surface has the correct trend with latitude, but shows a systematic deviation and fails to lie within the error bars.
Finally, the magnetic torque obtained from the model, $\Gamma_M = -0.549$, is the data which best predict the observed value $\Gamma_M = -0.553 \pm 0.014$. Its misfit goes down to $\chi = 0.52$.

\begin{figure}
 \includegraphics[width=\linewidth]{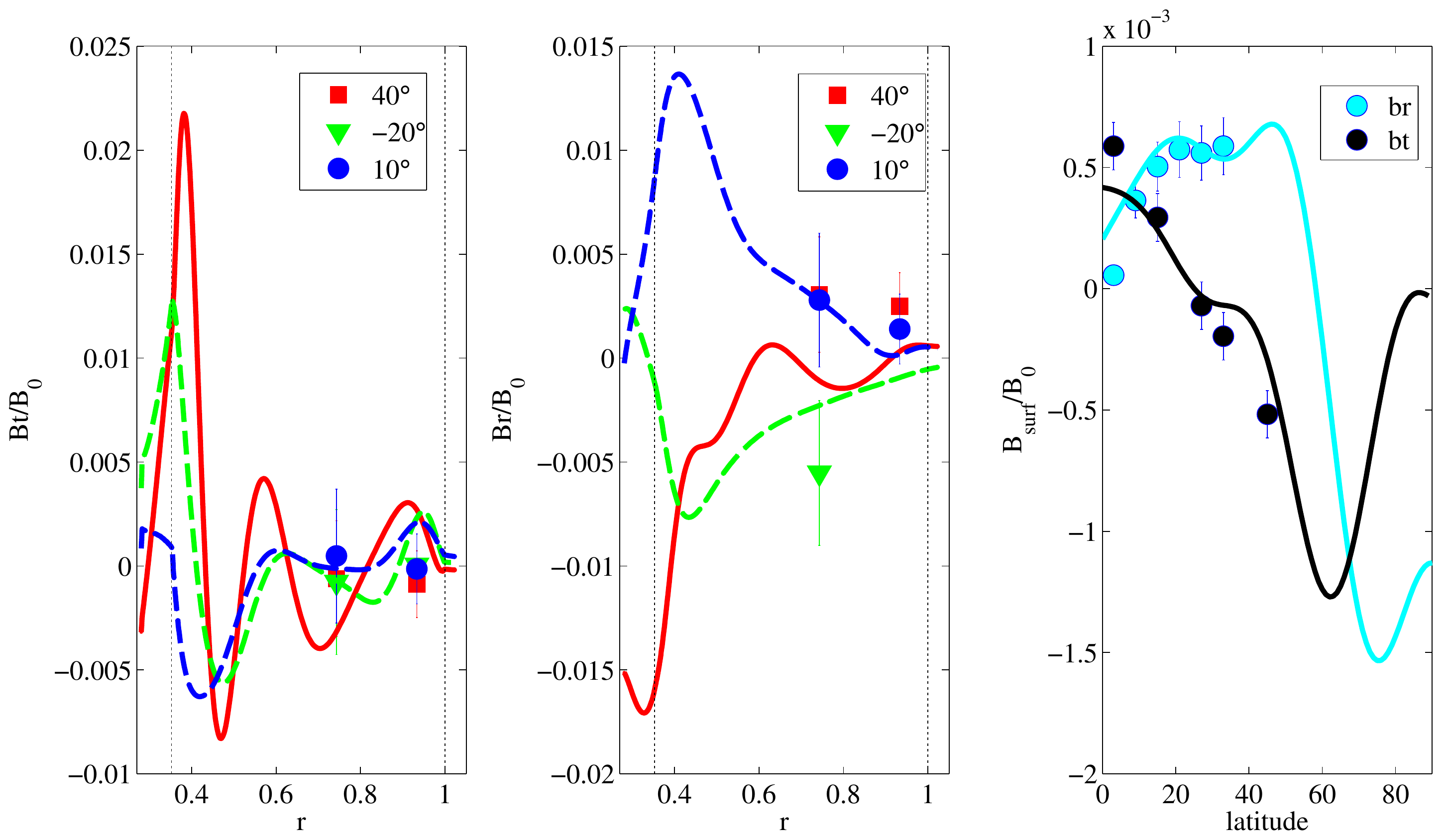}
       \caption{(Color online) Meridional predictions and measurements: these plots represent the measured data using marker symbols with their error bars and numerical predictions from our kinematic model using solid lines. The radial description of the magnetic field at three different latitudes ($40^{\circ}$ (squares and solid lines), $10^{\circ}$ (triangles and light gray dashed lines) and $-20^{\circ}$ (triangles and dark gray dashed lines))
is shown: the orthoradial component (left) and the radial component (center).
Radial (light gray) and orthoradial components (dark gray) of the induced magnetic field measured at the surface of the outer shell is also shown (right).
Radius ranges from $r_i-\delta_{Cu}$ to $r_o+\delta_{ss}$ and the vertical dotted lines indicate the fluid/solid interfaces at $r_i$ and $r_o$.
All fields are dimensionless (see Appendix \ref{Adimensionalization}).
\label{fits_merid_}}
\end{figure}

Figure \ref{fits_merid_} gives the fit to the radial and latitudinal components of the axisymmetric magnetic field, which is mainly due to the meridional circulation.
The model predicts reasonable amplitudes for both components.
Induced magnetic field inside the fluid shows radial oscillations, which are the direct consequence of the oscillations of the meridional velocity profiles (see Fig. \ref{Tp_c}).
This is especially true at high latitude where poloidal velocity is larger.
It is also remarkable that magnetic field on the surface is rather well predicted by our mean field kinematic model.
Finally, the global misfit of the whole data set, $\chi = 1.65$, constitutes a satisfactory result if we take into account that only the mean axisymmetric flow is used here.
Nevertheless, as it is slightly over unity, we argue that the effect of turbulent fluctuations could be considered to better fit magnetic induction.

\subsection{Non-axisymmetric data predictions\label{Non-axisymmetric data}}

\begin{figure}
\includegraphics[width=0.8\linewidth]{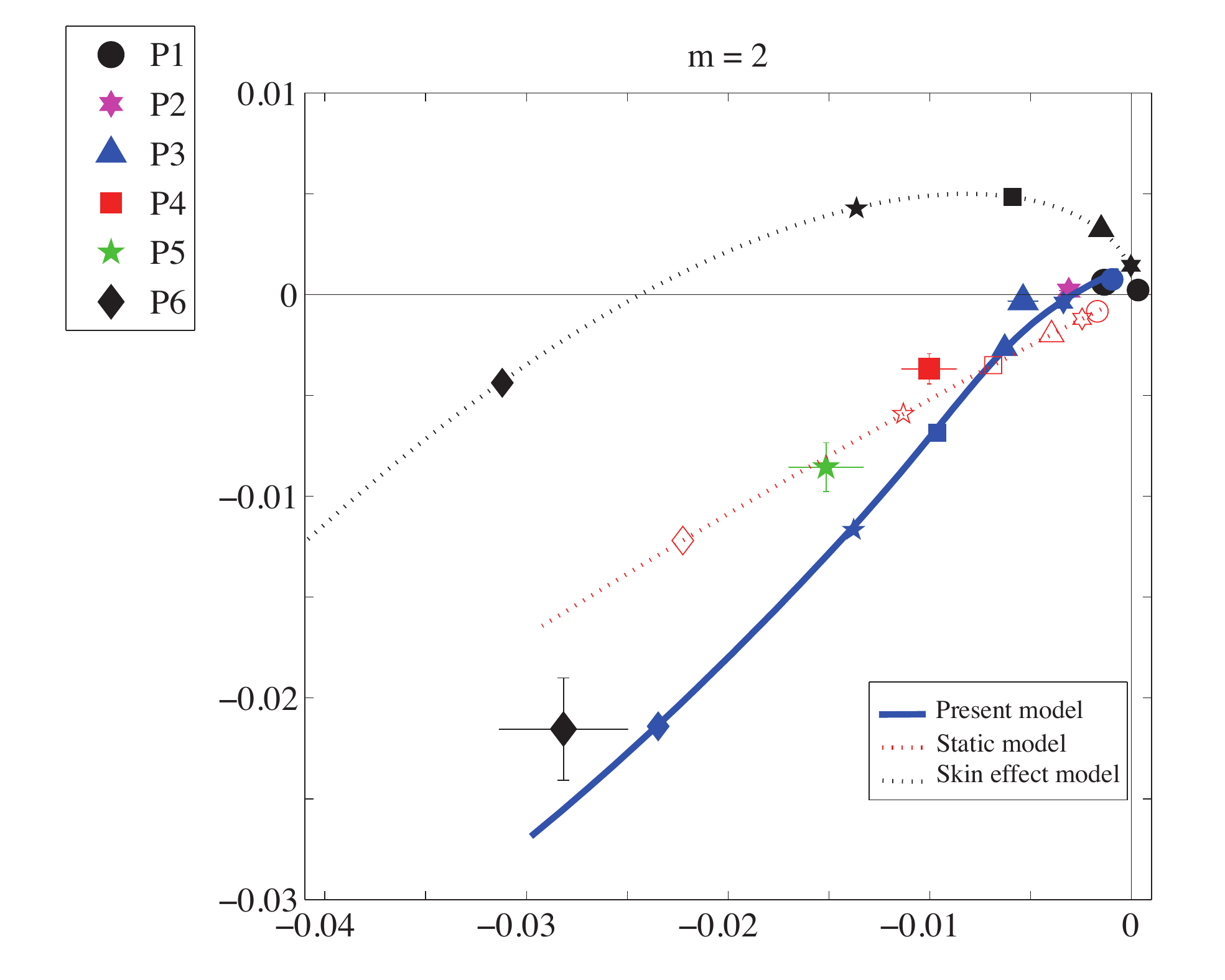}
        \caption{(Color online) Plot of the radial evolution of the non-axisymmetric magnetic data for experimental measurements and predictions at latitude $-10^{\circ}$. For experimental data, we plot the imaginary (vertical axis) and real part (horizontal axis) of the Fourier coefficients of the azimuthal motifs presented in section \ref{sub:motifs}.
Predictions display a full radial description, for three different models, presenting the real and imaginary part of $B^m_{\varphi}(r,\theta)$.
 We present on this graphic the azimuthal mode $m=2$ as an example.
Lines represent the radial evolution of $B^m_{\varphi}$ from the inner boundary to the outer sphere surface (where the magnetic field is numerically solved).
The thick solid blue (dark gray) line account for our \emph{kinematic} model predictions, the red dotted (bottom) line shows the \emph{static} case, and the black dotted (upper) line is the \emph{skin effect} case (see text sec. \ref{Non-axisymmetric data} for details).
Colored (grayscaled) markers labeled in the legend account for experimental data collected in DTS from magnetic probes in the sleeve.
Markers radially increase from P6 to P1 (see Figure \ref{Azim_pt}).
Non-axisymmetric data are all dimensionless (see Appendix \ref{Adimensionalization}).
\label{P_a10_m2}}
\end{figure}
We now comment how our model predicts the radial evolution of the non-axisymmetric magnetic field data, $\mathbf{B}(r,\theta)^m$ for each component and $m \ne 0$, when inner rotation rate is $f=-9$ Hz.
These data are computed as Fourier decomposition of $\mathbf{B}(r,\theta,\phi)$ and are comparable with Fourier coefficients presented in section \ref{non-axisymmetric data processing}.

Let's first focus on Figure \ref{P_a10_m2} representing the azimuthal component ($\varphi$) of non-axisymmetric mode $m=2$ at latitude $10^{\circ}$.  This graphic displays the radial description of the real and imaginary part of $B^m_{\varphi}(r,\theta)$ 
plotted in a complex plane.
Colored (grayscaled) markers, depicted in the caption, denote experimental observations at probes positions in the sleeve from the outermost probe $P1$ to the innermost $P6$.
They are reported in the complex plane with their real and imaginary error bars.

To improve our understanding of the physical processes involved in magnetic mode evolution, we show on Figure \ref{P_a10_m2} the result of three different models.
(i) Our \emph{kinematic} model predictions (as described in sec. \ref{profiles and predictions}) corresponds to the thick solid blue (dark gray) line where radial probe positions is reported using the same markers (colored in blue (dark gray)) as the one used for observations.
(ii) The \emph{static} case, where both inner and outer sphere are kept at rest, and the velocity field is zero everywhere, is represented by a red dashed (bottom) line.
In this model, each magnetic mode obeys the $r^{-(l+1)}$ law for a potential field, with a constant value imposed at the inner boundary ($r=r_i-\delta_{Cu}$).
Similarly, red (empty) markers indicate the static model predictions at radial probe positions.
(iii) The \emph{skin effect} case, where the non-axisymmetric boundary condition of the magnetic field $\mathbf{B}(t)$ is rotating and diffusing in a solid sodium shell (zero velocity field), is represented by the black dashed (upper) line.
Note that the copper layer has been replaced by an insulator (to prevent the shearing of magnetic field at the copper-sodium interface).
This simple \emph{skin effect} case, where no flow advection occurs, correspond to a purely diffusing model with a time-dependent external magnetic field, where equation \eqref{F_induction} is solved with $\mathbf{U_0} = 0$. Black (dark filled) markers also denote probe positions.

For all three models, we give in Fig. \ref{P_a10_m2} the radial evolution of the $m=2$ azimuthal magnetic mode from the inner boundary, near probe $P6$, to the outer surface close to probe $P1$.
Note that a straight line going through the origin means constant phase.
Radial amplitude variations are observed using marker locations and their distance to the complex plane origin.
It is also important to have in mind that probes from $P6$ to $P1$ are almost regularly spaced and that consequently constant marker spacing means linear amplitude variation.
Each model displays a radial evolution reflecting the physical processes involved, from a high amplitude value at the inner boundary (negative real and imaginary part in this case) to a low amplitude value at the outer boundary where magnetic mode amplitude get close to zero.
It is interesting to note that the different models give different magnetic field amplitude at the inner boundary.
This is a direct consequence of expression \eqref{bbc1} as the magnetic field at the inner boundary results from an inner source (imposed magnetic field) and an outer magnetic source (induced magnetic field). More details concerning boundary conditions are presented in section \ref{Magnetic boundary condition}.

Let's now describe the physical meaning of each model using the representation in a complex plane.
The \emph{static} case (bottom red dashed line) displays no phase shift all along its radial description. Its amplitude is rapidly decreasing from its innermost value (near probe $P6$) to probe $P1$.
This corresponds to the expected radial decrease proportional to $r^{-(l+1)}$ for $l=2$ to $11$ at $m=2$.
The time dependent \emph{skin effect} case (black dotted line) displays large phase variations, increasing radially and ending with a phase shift nearly equal to $\pi/2$ at the outer sphere.
This phase variation is due to electrical currents induced from the diffusion term of equation \ref{F_induction}.
Both the phase and the amplitude reflect this skin-depth effect.\\

Our velocity field model (thick solid line) provides a better fit to observations than the \emph{skin effect} and \emph{static} models.
We expected complex dynamics but surprisingly our model displays similarities with the two simple cases.
Indeed, the behavior of our kinematic model between probes $P6$ and $P4$ is very similar that of the static model: the phase is almost constant (straight line toward the complex plane origin) and the amplitude rapidly decrease with increasing radius.
This behavior is also observed experimentally for the same radius range (using markers positions).
An explanation is that in its inner region, DTS's flow is close to solid body rotation at the inner sphere spinning velocity (see section \ref{fields Maps} for spatial velocity description).
Magnetic field modes then behave as for the static case in the inner sphere rotating frame.
The flow dynamics is very different in the outer region. Velocity is decreasing and largely deviates from solid rotation.
We observe phase variations as for the \emph{skin effect} case from probe $P4$ to probe $P1$ and the very last probe $P1$ ends near the \emph{skin effect} model predictions. Observations display a similar trend and our model's predictions are not far from the measurements, although further away than permitted by the error bars.

\begin{figure}
 \includegraphics[width=\linewidth]{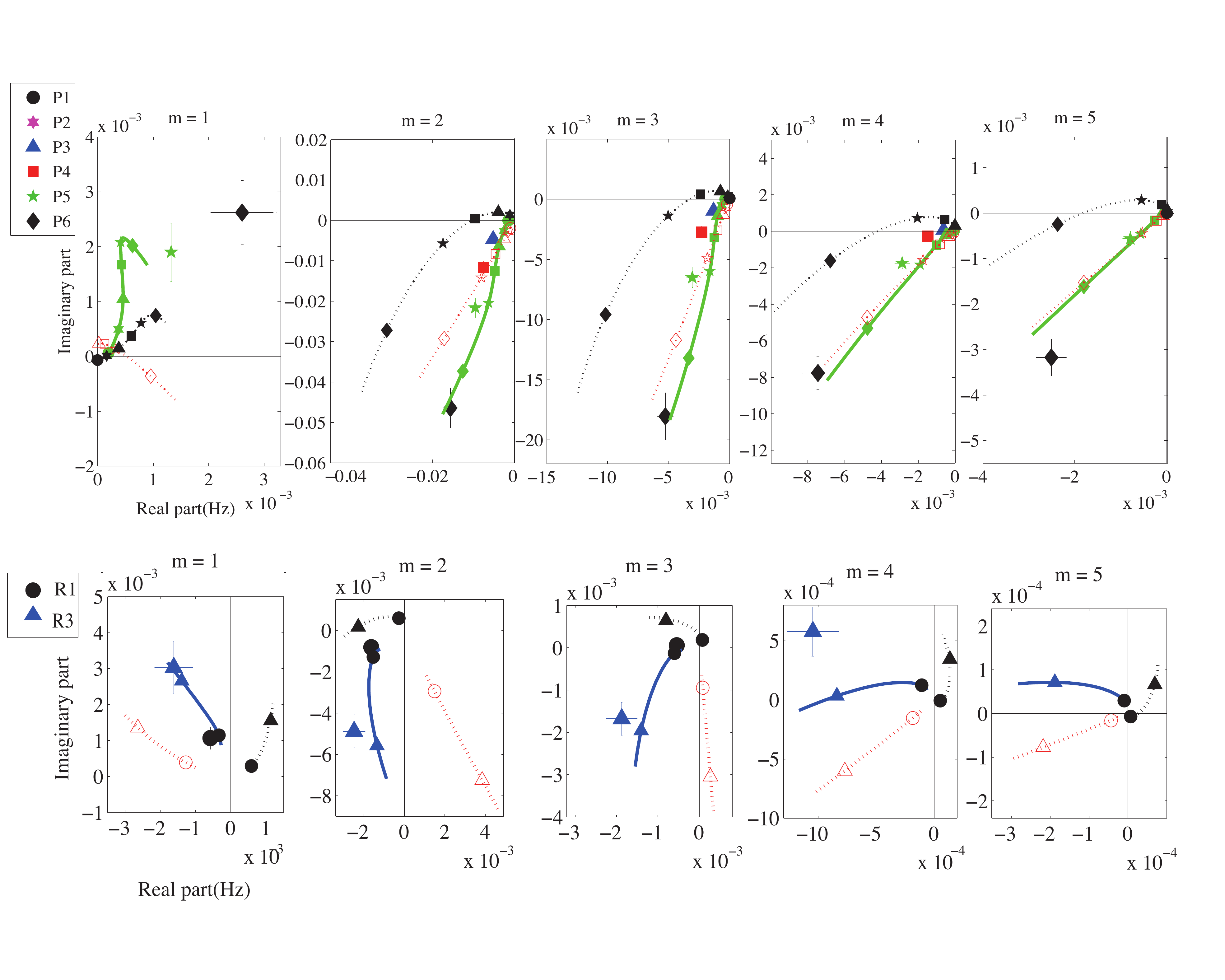}
  \caption{(Color online) Plot of non-axisymmetric magnetic data of five azimuthal modes from $m=1$ to $5$, for two magnetic components at two different latitudes.
   The azimuthal component $B_{\varphi}^m(r,\theta)$ is presented on the upper five graphics for latitude $-20^{\circ}$.
   The radial component $B_{r}^m(r,\theta)$ is presented on the bottom five graphics for latitude $10^{\circ}$.
   Colored (grayscaled) symbols labeled in the legend mark experimental data collected in DTS from magnetic probes in the sleeve.
   The radial position of the probes increases from P6 to P1 for the azimuthal component and from $R3$ to $R1$ for the radial component (see Figure \ref{Azim_pt}).
   Non axisymmetric data are dimensionless as given in Appendix \ref{Adimensionalization}.
   See caption of Figure \ref{P_a10_m2} for more details.
   \label{P_a-20mm}}
\end{figure}

Figure \ref{P_a-20mm} shows the radial ($r$) and azimuthal components ($\varphi$) of all the modes ($m=1$ to $m=5$) at two different latitudes ($-20^{\circ}$ and $10^{\circ}$).
One can observe that the azimuthal component of high order modes ($m=4$ and $5$) follow the trend of the \emph{static} case.
Our explanation is that high order modes are fed by high harmonic degrees which rapidly decrease in the fluid and reach a nearly zero amplitude in the outer flow region. This translates in the complex plane into the important spacing between probes $P6$ and $P5$ and to the outermost probes $P4$, $P3$, $P2$ and $P1$ gathering near the origin.
Lower order modes ($m=2$ and $3$) display a different behavior which deviates from the \emph{static} case as already discussed for $m=2$ in Figure \ref{P_a10_m2}.
Mode $m=1$ is closer to the \emph{skin effect} case, displaying no similarity with the \emph{static} case.
However, there are strong differences with the \emph{skin effect} case: the magnetic field is stretched to high amplitude values near the inner sphere.
The measurements are not well fit.

Radial component of the non-axisymmetric data 
lies at half way between \emph{skin effect} and \emph{static} cases.
We note however that high order modes get closer to the \emph{skin effect} model, while low order modes tend to recover a behavior closer to the \emph{static} case.

All those results show that low $m$ modes sound DTS's flow in its outer region while high $m$ modes collect information in the inner region.
It is of great interest that non-axisymmetric data are consistent with the predictions obtained from our velocity maps (Figure \ref{fields}).
It is likely that these data provide strong constraints in our inversion process and thus constitute a key ingredient for further investigations of turbulent fluctuations, as their misfit remains far from unity ($\chi=3.2$).

\section{Discussion\label{Discussions}}

We presented in this article a non linear inversion method to reconstruct the mean azimuthal and meridional velocity fields in a spherical liquid sodium device from measurements of the induced magnetic field, the electric potential and ultrasound Doppler velocity profiles.
To perform this inversion, we numerically compute the kinematic direct model solving the full induction equation for an axisymmetric flow and a time-dependent non-axisymmetric magnetic field.
Somehow the equation of motion is solved by the experiment and we compute a kinematic system governed by a single dimensionless number, the magnetic Reynolds number, which achieves the experimental value of $Rm = 28$. 

To solve this non linear kinematic system we used the code of \citet{figueroa13}.
We also had to take in consideration sharp conductivity jumps at the solid/fluid interfaces and realistic velocity boundary conditions.
We achieve remarkable predictions from a rather simple mean velocity model.
The best velocity field we obtain provides a full coherent solution for the mean velocity, the electric potential and the induced magnetic field inside and outside the fluid.
Indeed, it fits rather well the observations with a global normalized misfit of $1.65$.

Based on Nataf's previous investigation of DTS mean velocity field \cite{nataf13}, we extend the analysis to a higher $Rm$ value ($28$ instead of $9$).
The new velocity model displays a super-rotation zone, as the one described by \citet{nataf13} and \citet{brito11} at lower rotation rate, which confirms that super-rotation is confined to the innermost fluid region and that its relative strength slightly decreases when rotation rate increases.
On the contrary, the geostrophic shear region invades a large part of the fluid shell leading to an important toroidal induction by $\omega$-effect.
We expect this phenomenon to develop on a larger extent at even higher $Rm$, in such a way that the Coriolis force would increasingly dominate the dynamics.

The magnetostrophic regime also largely impacts the meridional flow behavior.
Recent results from  \citet{wicht14} show that when no magnetic field is imposed, the poloidal to toroidal kinetic energy ratio is about $18 \%$.
We now know from this study and the previous one \cite{nataf13} that we are far from a purely hydrodynamic system, as our strong imposed magnetic field largely damps the meridional flow.
In fact, the poloidal flow experiences a large Lorentz force by interacting with the magnetic field lines of the imposed dipole, which on the contrary acts as a propeller for the toroidal component.
As a consequence, the poloidal to toroidal kinetic energy ratio gets down to about $5 \%$ in the present model and to about $0.1\%$ in Nataf's model.
We partially attribute this difference between the two models to non-axisymmetric data, which require a larger meridional velocity.
We note that the previous model \cite{nataf13} at $Rm=9.4$ does not include such data.

Despite of its limited strength, the meridional circulation contributes to a significant part of the induced magnetic field.
In the mean field approach, the meriodional flow is the only way to feed the poloidal magnetic field and a slight meridional circulation interacting with a strong imposed magnetic field has an important induction effect.
In fact, poloidal to toroidal magnetic energy ratio jumps from $0.11$ in the previous model at $Rm=9.4$ \cite{nataf13} to $0.35$ in our model at $Rm=28$.
Meridional advection effects are also observed on the azimuthal magnetic field (see Figure \ref{fits_b}).
Estimating second order induction terms at $Rm=9.4$, \citet{nataf13}  came out with an amplitude deviation of the azimuthal induced magnetic field, due to meridional flow, to about $3 \%$ if we consider an induced magnetic field proportional to $Rm$ at first order.
The present reconstruction at $Rm=28$, involving larger poloidal kinetic energy, shows an increased local deviation of the azimuthal induction to about $5-10\%$.

Compared to previous work of \citet{nataf13}, we have introduced non-axisymmetric magnetic observations to obtain information on the mean axisymmetric flow.
Inspired by \citet{frick10}, we have developed a method to solve a time-varying magnetic field imposed from inner boundary and diffusing into a spherical Couette flow.
Prediction results and observations of the non-axisymmetric part of the magnetic field are compared using a complex plane representation describing the radial evolution of each non-axisymmetric magnetic mode.
By comparison with two reference cases (\emph{static} and \emph{skin effect}, see sec. \ref{Non-axisymmetric data}) we evidence two dominant dynamics that split the DTS flow in a solid body rotation part near the inner-core and a shearing zone near the outer shell.
This picture is coherent with the flow structure observed on velocity maps.
The next step is to make use of our new inversion procedure to investigate the effect of turbulent fluctuations at larger magnetic Reynolds number.

We used a realistic induction model which takes into account both meridional and azimuthal velocity fields, and invert for the mean veolocity field that best explains our measurements in a magnetostrophic regime.
We thus bridged the gap to investigate non-axisymmetric fluctuations which are thought to play a major role in the generation of a large scale magnetic field in dynamo theory.
A vast literature exists on mean magnetic field generation by a mean electromotive force $\epsilon = \langle \tilde u \times \tilde b\rangle$ induced by fluctuating non-axisymmetric flow $\tilde u$ and magnetic field $\tilde b$, acting as an additional source term in the induction equation \eqref{F_induction}.
A usual approach is to make use of the mean electromotive force expansion in terms of the large-scale magnetic field and the $\alpha$ and $\beta$ tensors, as presented in Sec. \ref{Introduction}.
\citep{Cabanes14b} follow this strategy and invert for both the mean velocity field and radial profiles of $\alpha$ and $\beta$ (assumed to
be scalar).



\appendix

\section{Adimensionalization\label{Adimensionalization}}
 
See table \ref{tab:adimentionalization}.
 
\begin{table}
\caption{Scales used in this paper for adimensionalization.
 $\Omega$ is the imposed angular velocity of the inner sphere and $r_o^{\ast}$ is the dimensional inner radius of the outer shell.
$B_o$ is the intensity of the dipolar magnetic field at the equator of the outer shell at radius $r=r_o$.
Rm is the magnetic Reynolds number defined in table \ref{tab:dimensionless numbers}.
We call $\eta$ the magnetic diffusivity, $\rho$ the density, and $\mu$ the magnetic permittivity of liquid sodium.
\label{tab:adimentionalization}}
\begin{ruledtabular}
\begin{tabular}{ccccccccc}
{Quantities} & {scale} \\
\colrule
{time} & $\Omega^{-1}$\\
{length} & $r_o^{\ast}$\\
{velocity} & $r_o^{\ast} \Omega$\\
{imposed and non-axisymmetric magnetic field} & $B_o$\\
{scalar magnetic potential} & $r_o^{\ast} B_o$\\
{induced axisymmetric magnetic field} & $Rm B_o$\\
{electric potential} & $Rm \eta B_o$\\
{magnetic torque} & $Rm (r_o^{\ast})^3 B_o^2/ \mu $\\
{energy} & $\rho \Omega^2(r_o^{\ast})^5 = \rho \eta^2 r_o^{\ast} Rm^2$\\
\end{tabular}
 \end{ruledtabular}
 \end{table}
 
\section{inversion of the magnetic potential of the magnet}
\label{magnetic potential}

We determine the magnetic potential of the magnet of the inner sphere at rest and in the absence of the sodium layer, by inverting measurements of components of the magnetic field taken in the gap between the inner sphere and the outer sphere and at the surface of the latter.

\begin{figure}
\includegraphics[width=0.5\linewidth]{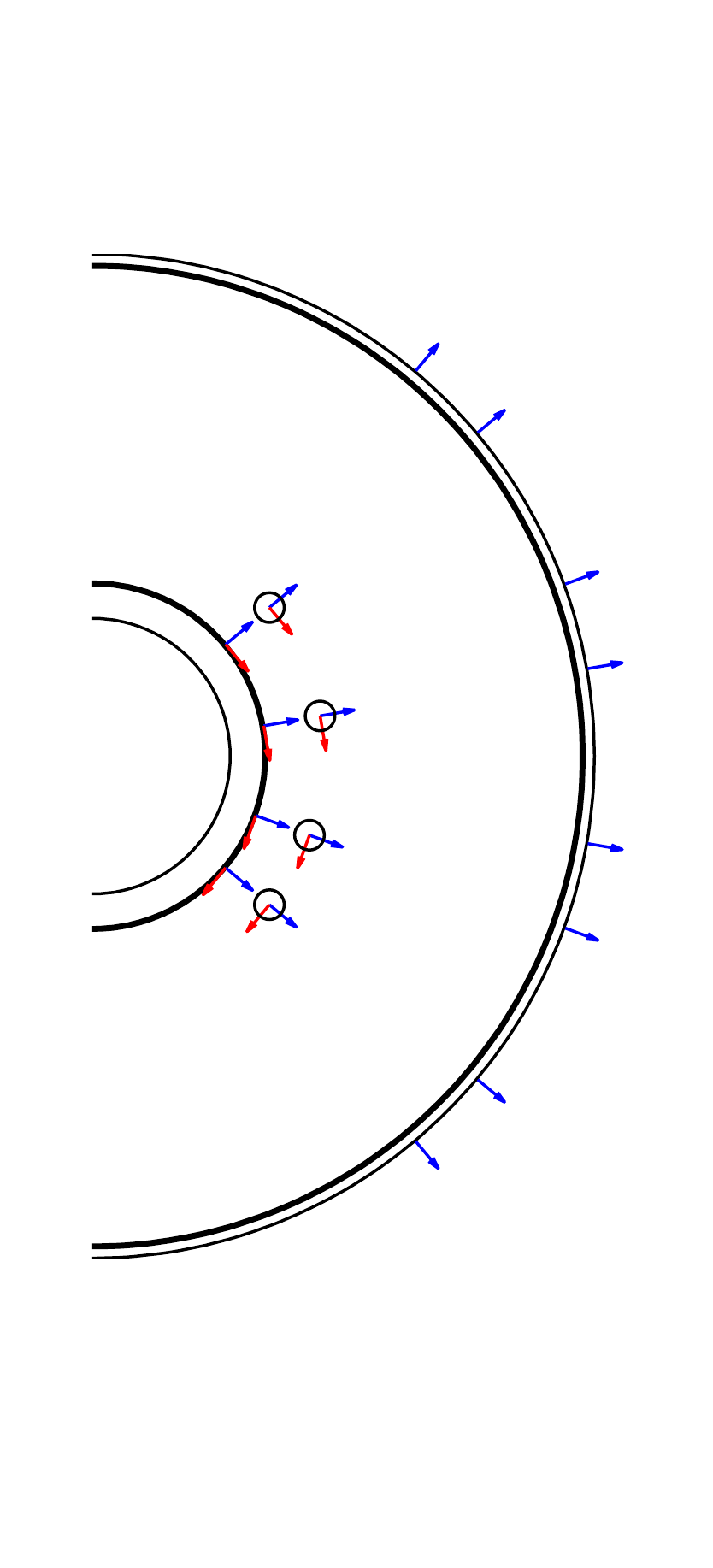}
\caption{(Color online) Location of the magnetic measurements used to infer the scalar magnetic potential of the inner sphere's magnet.
In this meridional cross-section, the arrows indicate which component ($B_r$ in blue, $B_{\theta}$ in red) is measured, the small black circles corresponding to $B_\varphi$ data.
The measurements are repeated every $5^\circ$ or $10^\circ$ in longitude.
Black spherical lines mark successively the copper, the fluid and the stainless steel shells.
\label{graine_measurement_points}}
\end{figure}

\begin{figure}
 \centering
 \includegraphics[width=\linewidth]{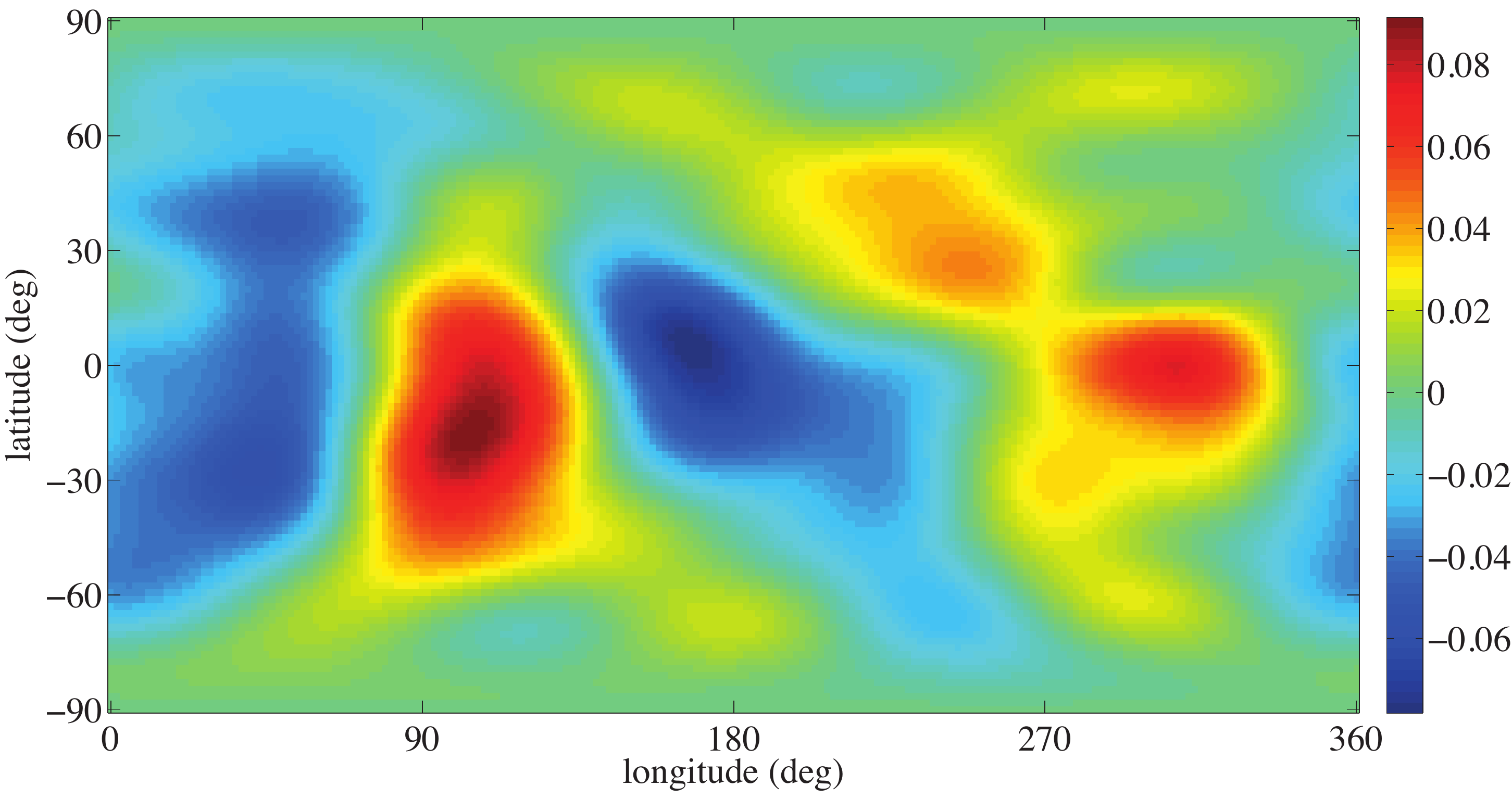}
 \includegraphics[width=\linewidth]{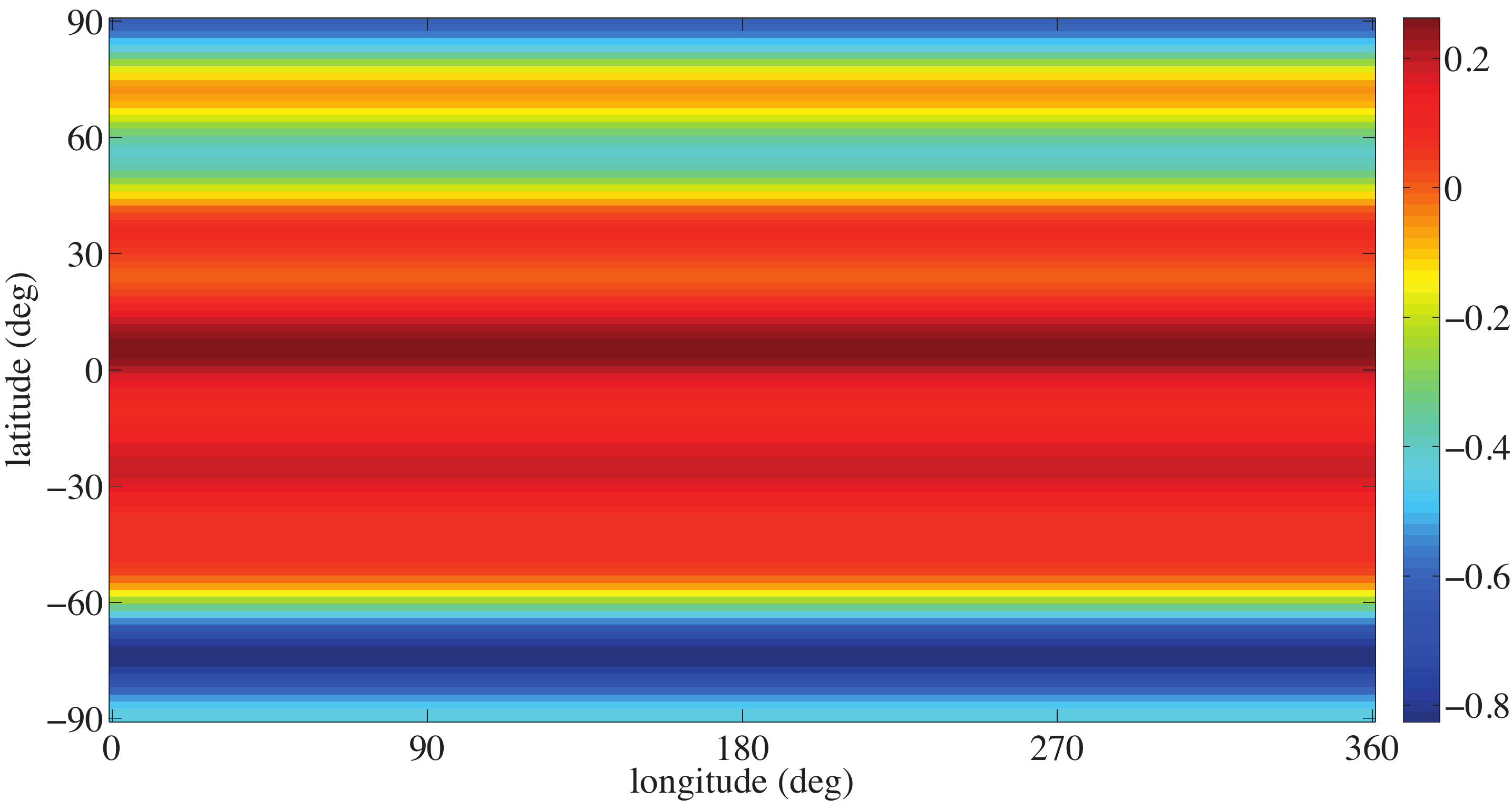}
\caption{(Color online) Contour map of the scalar magnetic potential of the magnet reconstructed at the surface of the inner sphere.
The potential is normalized by $r_o^{\ast} B_o$ as indicated in table \ref{tab:adimentionalization}.
{\it Top:} non-axisymmetric part.  
{\it Bottom:} non-dipole axisymmetric part.  
\label{dual}}
\end{figure}

We measure $B_r$, $B_\theta$ and $B_\varphi$ and their standard deviation, using a Hirst GM05 magnetometer.
Figure \ref{graine_measurement_points} shows the location of the measurements in a meridional plane.
The measurements are repeated every $5^\circ$ or $10^\circ$ in longitude, yielding a total of $1296$ data points.
We then invert for the magnetic scalar potential $A(r,\theta,\varphi)$ at the surface of the inner sphere, using:
\begin{equation}
\boldsymbol{B} = -\boldsymbol{\nabla}A.
\end{equation}
The scalar potential is projected on spherical harmonics:
\begin{equation}
A(r,\theta,\varphi) = r_i \sum_{l=1}^{L} \left( \frac{r_i}{r} \right)^{l+1} \sum_{m=0}^{min(l,M)} \left[ g_l^m \cos m \varphi + h_l^m \sin m \varphi \right] P_l^m(\cos\theta)\, ,
\end{equation}
and we invert for the spherical harmonics coefficients up to degree $L=11$ and order $M=6$.
For this over-constrained inversion, we use the {\it l1qc\_logbarrier} $l1$-norm algorithm of \citet{candes05} in order to minimize the number of non-zero coefficients and avoid the blow-up of the potential at high latitudes.
Figure \ref{dual} displays contour maps of the non-axisymmetric part ($m\ne0$) and axisymmetric part ($m=0$) of our preferred model of scalar magnetic potential at the surface of the inner sphere.
For this model, the normalized misfit of the non-axisymmetric components is $0.69$.
The motif shown in Figure \ref{motif} for $f=0$ is computed from this potential.

\section{Finite difference formulation for conductivity jumps\label{fd_jumps}}

For discontinuous conductivities (a solid boundary with a different conductivity), the magnetic field $\mathbf{B}$ is continuous across the interface, which translates into the poloidal-toroidal formalism into continuous $B_P$, $\partial_r B_P$ and $B_T$.
In addition, 
The second jump condition is due to the continuity of the tangential electric field, leadting to $(j \times \mathbf{n})/\sigma$ continuous across the interface, from which we have the continuity of $\eta \partial_r (rB_T)$ and $\eta \Delta B_P$ (if magnetic permeability $\mu_0$ is the same everywhere).

Since $\eta$ is constant everywhere but at the discontinuity, the governing equation is
\begin{equation}
\partial_t \mathbf{b} = \mathbf{\nabla} \times ( \mathbf{u} \times \mathbf{b} ) + \eta \Delta \mathbf{b}
\end{equation}

In order to derive the finite difference approximation, we write the Taylor expansions left and right of the discontinuity located at $r_0$:
\begin{equation}
f_\pm = f_{0\pm} \pm dr (\partial_r f)_{0\pm}  +  \frac{dr^2}{2} (\partial_{rr} f)_{0\pm} \label{eq:f_pm}
\end{equation}
where $f_\pm = f(r_0 \pm dr)$, and $f_{0\pm}$ is the limit $f(r_0 \pm \varepsilon)$ when $\varepsilon \to 0^+$.
This is the start for a finite difference formulation, with equidistant points around the discontinuity.

We can isolate $\partial_{rr} f$
\begin{equation}
2\frac{f_\pm - f_{0\pm}}{dr^2} \mp \frac{2}{dr} (\partial_r f)_{0\pm} = (\partial_{rr} f)_{0\pm}
\end{equation}
and add the angular part to have an evaluation of $\Delta (f/r)$ at each side of the discontinuity:
\begin{equation}
2\frac{f_\pm - f_{0\pm}}{dr^2} \mp \frac{2}{dr} (\partial_r f)_{0\pm} - \frac{\ell(\ell+1)}{r_0^2} f_{0\pm}
	= (r\Delta (f/r))_{0\pm}
	\label{eq:fd_jump}
\end{equation}

\subsubsection{Poloidal part}

Here, we take $f = rP$ in the previous finite difference formulation.
In this case, $f_0$ and $(\partial_r f)_0$ are uniquely defined (by continuity).
Multiplying equation \ref{eq:fd_jump} by $\eta_+ \eta_-$ gives:
\begin{equation}
2\eta_+ \eta_- \left( \frac{ f_\pm - f_0 }{dr^2} - \frac{\ell(\ell+1)}{2r_0^2} f_0 \right)= \pm \eta_+ \eta_- \frac{2}{dr} (\partial_r f)_0 + \eta_+ \eta_- (r\Delta P)_{0\pm}
\end{equation}
By taking the sum of these equations, we eliminate $(\partial_r f)_0$:
\begin{equation}
2\eta_+ \eta_- \left( \frac{ f_+ + f_- - 2f_0 }{dr^2} - \frac{\ell(\ell+1)}{r_0^2} f_0 \right) = (\eta_+ + \eta_-) r_0 K
\end{equation}
with $K = \eta_+ (\Delta P)_+ = \eta_- (\Delta P)_-$.
We finally obtain the finite difference approximation of $\eta \Delta P$:
\begin{equation}
(\eta \Delta P)_0 = \frac{2\eta_+ \eta_-}{\eta_+ + \eta_-} \left( \frac{ r_+ P_+ + r_- P_- - 2r_0 P_0}{r_0 dr^2} - \frac{\ell(\ell+1)}{r_0^2} \right)
	\label{eq:dP_jump}
\end{equation}
Meaning that, for the finite difference formulation, $\eta$ must simply be replaced by its harmonic mean at the interface. Or equivalently $\sigma$ be replaced by its mean.

\subsubsection{Toroidal part}

Here, we take $f = rT$ in the finite difference formulation of equations \ref{eq:f_pm}.
In this case, $f_0$ is uniquely defined (by continuity) and $\eta_+ (\partial_r f)_{0+} = \eta_- (\partial_r f)_{0-}$.
Multiplying equation \ref{eq:fd_jump} by $\eta_\pm$ gives:
\begin{equation}
2\eta_\pm \left( \frac{ f_\pm - f_0 }{dr^2} - \frac{\ell(\ell+1)}{2r_0^2} f_0 \right)= \pm \eta_\pm \frac{2}{dr} (\partial_r f)_{0\pm} + \eta_\pm (r\Delta T)_{0\pm}
\end{equation}
By taking the sum of these equations, we eliminate $\eta_\pm (\partial_r f)_{0\pm}$:
\begin{equation}
2\frac{\eta_+ f_+ + \eta_- f_- - (\eta_+ + \eta_-)f_0 }{dr^2} - (\eta_+ + \eta_-) \frac{\ell(\ell+1)}{r_0^2} f_0 = 2 r_0 K
\end{equation}
with $K = \eta_+ (\Delta T)_+ = \eta_- (\Delta T)_-$.
We finally obtain the finite difference approximation of $\eta \Delta T$:
\begin{equation}
(\eta \Delta T)_0 = \frac{\eta_+ r_+ T_+ + \eta_- r_- T_- - (\eta_+ + \eta_-) r_0 T_0 }{r_0 dr^2}
	- \frac{\eta_+ + \eta_-}{2} \frac{\ell(\ell+1)}{r_0^2} T_0
	\label{eq:dT_jump}
\end{equation}

\bibliography{biblio_simon_300714}

\begin{thebibliography}{30}%
\makeatletter
\providecommand \@ifxundefined [1]{%
 \@ifx{#1\undefined}
}%
\providecommand \@ifnum [1]{%
 \ifnum #1\expandafter \@firstoftwo
 \else \expandafter \@secondoftwo
 \fi
}%
\providecommand \@ifx [1]{%
 \ifx #1\expandafter \@firstoftwo
 \else \expandafter \@secondoftwo
 \fi
}%
\providecommand \natexlab [1]{#1}%
\providecommand \enquote  [1]{``#1''}%
\providecommand \bibnamefont  [1]{#1}%
\providecommand \bibfnamefont [1]{#1}%
\providecommand \citenamefont [1]{#1}%
\providecommand \href@noop [0]{\@secondoftwo}%
\providecommand \href [0]{\begingroup \@sanitize@url \@href}%
\providecommand \@href[1]{\@@startlink{#1}\@@href}%
\providecommand \@@href[1]{\endgroup#1\@@endlink}%
\providecommand \@sanitize@url [0]{\catcode `\\12\catcode `\$12\catcode
  `\&12\catcode `\#12\catcode `\^12\catcode `\_12\catcode `\%12\relax}%
\providecommand \@@startlink[1]{}%
\providecommand \@@endlink[0]{}%
\providecommand \url  [0]{\begingroup\@sanitize@url \@url }%
\providecommand \@url [1]{\endgroup\@href {#1}{\urlprefix }}%
\providecommand \urlprefix  [0]{URL }%
\providecommand \Eprint [0]{\href }%
\providecommand \doibase [0]{http://dx.doi.org/}%
\providecommand \selectlanguage [0]{\@gobble}%
\providecommand \bibinfo  [0]{\@secondoftwo}%
\providecommand \bibfield  [0]{\@secondoftwo}%
\providecommand \translation [1]{[#1]}%
\providecommand \BibitemOpen [0]{}%
\providecommand \bibitemStop [0]{}%
\providecommand \bibitemNoStop [0]{.\EOS\space}%
\providecommand \EOS [0]{\spacefactor3000\relax}%
\providecommand \BibitemShut  [1]{\csname bibitem#1\endcsname}%
\let\auto@bib@innerbib\@empty
\bibitem [{\citenamefont {Cowling}(1933)}]{cowling33}%
  \BibitemOpen
  \bibfield  {author} {\bibinfo {author} {\bibfnamefont {T.}~\bibnamefont
  {Cowling}},\ }\href {\doibase 10.1093/mnras/94.1.39} {\bibfield  {journal}
  {\bibinfo  {journal} {Mon. Not. Roy. Astron. Soc.}\ }\textbf {\bibinfo
  {volume} {94}},\ \bibinfo {pages} {39} (\bibinfo {year} {1933})}\BibitemShut
  {NoStop}%
\bibitem [{\citenamefont {Ponomarenko}(1973)}]{ponomarenko73}%
  \BibitemOpen
  \bibfield  {author} {\bibinfo {author} {\bibfnamefont {Y.}~\bibnamefont
  {Ponomarenko}},\ }\href {\doibase 10.1007/BF00853190} {\bibfield  {journal}
  {\bibinfo  {journal} {J. Appl. Mech. Tech. Phys.}\ }\textbf {\bibinfo
  {volume} {14}},\ \bibinfo {pages} {775} (\bibinfo {year} {1973})}\BibitemShut
  {NoStop}%
\bibitem [{\citenamefont {{Gailitis}}\ \emph {et~al.}(2001)\citenamefont
  {{Gailitis}}, \citenamefont {{Lielausis}}, \citenamefont {{Platacis}},
  \citenamefont {{Dement'ev}}, \citenamefont {{Cifersons}}, \citenamefont
  {{Gerbeth}}, \citenamefont {{Gundrum}}, \citenamefont {{Stefani}},
  \citenamefont {{Christen}},\ and\ \citenamefont {{Will}}}]{gailitis01}%
  \BibitemOpen
  \bibfield  {author} {\bibinfo {author} {\bibfnamefont {A.}~\bibnamefont
  {{Gailitis}}}, \bibinfo {author} {\bibfnamefont {O.}~\bibnamefont
  {{Lielausis}}}, \bibinfo {author} {\bibfnamefont {E.}~\bibnamefont
  {{Platacis}}}, \bibinfo {author} {\bibfnamefont {S.}~\bibnamefont
  {{Dement'ev}}}, \bibinfo {author} {\bibfnamefont {A.}~\bibnamefont
  {{Cifersons}}}, \bibinfo {author} {\bibfnamefont {G.}~\bibnamefont
  {{Gerbeth}}}, \bibinfo {author} {\bibfnamefont {T.}~\bibnamefont
  {{Gundrum}}}, \bibinfo {author} {\bibfnamefont {F.}~\bibnamefont
  {{Stefani}}}, \bibinfo {author} {\bibfnamefont {M.}~\bibnamefont
  {{Christen}}}, \ and\ \bibinfo {author} {\bibfnamefont {G.}~\bibnamefont
  {{Will}}},\ }\href {\doibase 10.1103/PhysRevLett.86.3024} {\bibfield
  {journal} {\bibinfo  {journal} {Phys. Rev. Lett.}\ }\textbf {\bibinfo
  {volume} {86}},\ \bibinfo {pages} {3024} (\bibinfo {year}
  {2001})}\BibitemShut {NoStop}%
\bibitem [{\citenamefont {Charbonneau}(2010)}]{charbonneau10}%
  \BibitemOpen
  \bibfield  {author} {\bibinfo {author} {\bibfnamefont {P.}~\bibnamefont
  {Charbonneau}},\ }\href {\doibase 10.12942/lrsp-2010-3} {\bibfield  {journal}
  {\bibinfo  {journal} {Living Reviews in Solar Physics}\ }\textbf {\bibinfo
  {volume} {7}} (\bibinfo {year} {2010}),\ 10.12942/lrsp-2010-3}\BibitemShut
  {NoStop}%
\bibitem [{\citenamefont {Steenbeck}\ \emph {et~al.}(1966)\citenamefont
  {Steenbeck}, \citenamefont {Krause},\ and\ \citenamefont
  {R{\"a}dler}}]{steenbeck66}%
  \BibitemOpen
  \bibfield  {author} {\bibinfo {author} {\bibfnamefont {M.}~\bibnamefont
  {Steenbeck}}, \bibinfo {author} {\bibfnamefont {F.}~\bibnamefont {Krause}}, \
  and\ \bibinfo {author} {\bibfnamefont {K.}~\bibnamefont {R{\"a}dler}},\
  }\href@noop {} {\bibfield  {journal} {\bibinfo  {journal} {Z. Naturforsch}\ }
  (\bibinfo {year} {1966})}\BibitemShut {NoStop}%
\bibitem [{\citenamefont {Krause}\ and\ \citenamefont
  {R{\"a}dler}(1980)}]{krause80}%
  \BibitemOpen
  \bibfield  {author} {\bibinfo {author} {\bibfnamefont {F.}~\bibnamefont
  {Krause}}\ and\ \bibinfo {author} {\bibfnamefont {K.-H.}\ \bibnamefont
  {R{\"a}dler}},\ }\href {http://tocs.ulb.tu-darmstadt.de/80642187.pdf} {\emph
  {\bibinfo {title} {Mean-field magnetohydrodynamics and dynamo theory}}}\
  (\bibinfo  {publisher} {Pergamon Press Oxford},\ \bibinfo {year}
  {1980})\BibitemShut {NoStop}%
\bibitem [{\citenamefont {R{\"u}diger}\ and\ \citenamefont
  {Brandenburg}(2014)}]{rudiger14}%
  \BibitemOpen
  \bibfield  {author} {\bibinfo {author} {\bibfnamefont {G.}~\bibnamefont
  {R{\"u}diger}}\ and\ \bibinfo {author} {\bibfnamefont {A.}~\bibnamefont
  {Brandenburg}},\ }\href {\doibase 10.1103/PhysRevE.89.033009} {\bibfield
  {journal} {\bibinfo  {journal} {Phys. Rev. E}\ }\textbf {\bibinfo {volume}
  {89}},\ \bibinfo {pages} {033009} (\bibinfo {year} {2014})}\BibitemShut
  {NoStop}%
\bibitem [{\citenamefont {Frick}\ \emph {et~al.}(2010)\citenamefont {Frick},
  \citenamefont {Noskov}, \citenamefont {Denisov},\ and\ \citenamefont
  {Stepanov}}]{frick10}%
  \BibitemOpen
  \bibfield  {author} {\bibinfo {author} {\bibfnamefont {P.}~\bibnamefont
  {Frick}}, \bibinfo {author} {\bibfnamefont {V.}~\bibnamefont {Noskov}},
  \bibinfo {author} {\bibfnamefont {S.}~\bibnamefont {Denisov}}, \ and\
  \bibinfo {author} {\bibfnamefont {R.}~\bibnamefont {Stepanov}},\ }\href
  {http://dx.doi.org/10.1103/PhysRevLett.105.184502} {\bibfield  {journal}
  {\bibinfo  {journal} {Phys. Rev. Lett.}\ }\textbf {\bibinfo {volume} {105}},\
  \bibinfo {pages} {184502} (\bibinfo {year} {2010})}\BibitemShut {NoStop}%
\bibitem [{\citenamefont {Rahbarnia}\ \emph {et~al.}(2012)\citenamefont
  {Rahbarnia}, \citenamefont {Brown}, \citenamefont {Clark}, \citenamefont
  {Kaplan}, \citenamefont {Nornberg}, \citenamefont {Rasmus}, \citenamefont
  {Zane~Taylor}, \citenamefont {Forest}, \citenamefont {Jenko}, \citenamefont
  {Limone}, \citenamefont {Pinton}, \citenamefont {Plihon},\ and\ \citenamefont
  {Verhille}}]{rahbarnia12}%
  \BibitemOpen
  \bibfield  {author} {\bibinfo {author} {\bibfnamefont {K.}~\bibnamefont
  {Rahbarnia}}, \bibinfo {author} {\bibfnamefont {B.~P.}\ \bibnamefont
  {Brown}}, \bibinfo {author} {\bibfnamefont {M.~M.}\ \bibnamefont {Clark}},
  \bibinfo {author} {\bibfnamefont {E.~J.}\ \bibnamefont {Kaplan}}, \bibinfo
  {author} {\bibfnamefont {M.~D.}\ \bibnamefont {Nornberg}}, \bibinfo {author}
  {\bibfnamefont {A.~M.}\ \bibnamefont {Rasmus}}, \bibinfo {author}
  {\bibfnamefont {N.}~\bibnamefont {Zane~Taylor}}, \bibinfo {author}
  {\bibfnamefont {C.~B.}\ \bibnamefont {Forest}}, \bibinfo {author}
  {\bibfnamefont {F.}~\bibnamefont {Jenko}}, \bibinfo {author} {\bibfnamefont
  {A.}~\bibnamefont {Limone}}, \bibinfo {author} {\bibfnamefont {J.-F.}\
  \bibnamefont {Pinton}}, \bibinfo {author} {\bibfnamefont {N.}~\bibnamefont
  {Plihon}}, \ and\ \bibinfo {author} {\bibfnamefont {G.}~\bibnamefont
  {Verhille}},\ }\href {http://dx.doi.org/10.1088/0004-637X/759/2/80}
  {\bibfield  {journal} {\bibinfo  {journal} {{Astrophys. J.}}\ }\textbf
  {\bibinfo {volume} {759}},\ \bibinfo {pages} {80} (\bibinfo {year}
  {2012})}\BibitemShut {NoStop}%
\bibitem [{\citenamefont {{Stieglitz}}\ and\ \citenamefont
  {{M\"uller}}(2001)}]{stieglitz01}%
  \BibitemOpen
  \bibfield  {author} {\bibinfo {author} {\bibfnamefont {R.}~\bibnamefont
  {{Stieglitz}}}\ and\ \bibinfo {author} {\bibfnamefont {U.}~\bibnamefont
  {{M\"uller}}},\ }\href {\doibase 10.1063/1.1331315} {\bibfield  {journal}
  {\bibinfo  {journal} {Phys. Fluids}\ }\textbf {\bibinfo {volume} {13}},\
  \bibinfo {pages} {561} (\bibinfo {year} {2001})}\BibitemShut {NoStop}%
\bibitem [{\citenamefont {{Roberts}}(1972)}]{roberts72}%
  \BibitemOpen
  \bibfield  {author} {\bibinfo {author} {\bibfnamefont {G.~O.}\ \bibnamefont
  {{Roberts}}},\ }\href {\doibase 10.1098/rsta.1972.0015} {\bibfield  {journal}
  {\bibinfo  {journal} {Phil. Trans. R. Soc. London, Ser. A}\ }\textbf
  {\bibinfo {volume} {271}},\ \bibinfo {pages} {411} (\bibinfo {year}
  {1972})}\BibitemShut {NoStop}%
\bibitem [{\citenamefont {Colgate}\ \emph {et~al.}(2011)\citenamefont
  {Colgate}, \citenamefont {Beckley}, \citenamefont {Si}, \citenamefont
  {Martinic}, \citenamefont {Westpfahl}, \citenamefont {Slutz}, \citenamefont
  {Westrom}, \citenamefont {Klein}, \citenamefont {Schendel}, \citenamefont
  {Scharle} \emph {et~al.}}]{colgate11}%
  \BibitemOpen
  \bibfield  {author} {\bibinfo {author} {\bibfnamefont {S.~A.}\ \bibnamefont
  {Colgate}}, \bibinfo {author} {\bibfnamefont {H.}~\bibnamefont {Beckley}},
  \bibinfo {author} {\bibfnamefont {J.}~\bibnamefont {Si}}, \bibinfo {author}
  {\bibfnamefont {J.}~\bibnamefont {Martinic}}, \bibinfo {author}
  {\bibfnamefont {D.}~\bibnamefont {Westpfahl}}, \bibinfo {author}
  {\bibfnamefont {J.}~\bibnamefont {Slutz}}, \bibinfo {author} {\bibfnamefont
  {C.}~\bibnamefont {Westrom}}, \bibinfo {author} {\bibfnamefont
  {B.}~\bibnamefont {Klein}}, \bibinfo {author} {\bibfnamefont
  {P.}~\bibnamefont {Schendel}}, \bibinfo {author} {\bibfnamefont
  {C.}~\bibnamefont {Scharle}},  \emph {et~al.},\ }\href {\doibase
  10.1103/PhysRevLett.106.175003} {\bibfield  {journal} {\bibinfo  {journal}
  {Phys. Rev. Lett.}\ }\textbf {\bibinfo {volume} {106}},\ \bibinfo {pages}
  {175003} (\bibinfo {year} {2011})}\BibitemShut {NoStop}%
\bibitem [{\citenamefont {Brito}\ \emph {et~al.}(2011)\citenamefont {Brito},
  \citenamefont {Alboussiere}, \citenamefont {Cardin}, \citenamefont
  {Gagni{\`e}re}, \citenamefont {Jault}, \citenamefont {La~Rizza},
  \citenamefont {Masson}, \citenamefont {Nataf},\ and\ \citenamefont
  {Schmitt}}]{brito11}%
  \BibitemOpen
  \bibfield  {author} {\bibinfo {author} {\bibfnamefont {D.}~\bibnamefont
  {Brito}}, \bibinfo {author} {\bibfnamefont {T.}~\bibnamefont {Alboussiere}},
  \bibinfo {author} {\bibfnamefont {P.}~\bibnamefont {Cardin}}, \bibinfo
  {author} {\bibfnamefont {N.}~\bibnamefont {Gagni{\`e}re}}, \bibinfo {author}
  {\bibfnamefont {D.}~\bibnamefont {Jault}}, \bibinfo {author} {\bibfnamefont
  {P.}~\bibnamefont {La~Rizza}}, \bibinfo {author} {\bibfnamefont {J.-P.}\
  \bibnamefont {Masson}}, \bibinfo {author} {\bibfnamefont {H.-C.}\
  \bibnamefont {Nataf}}, \ and\ \bibinfo {author} {\bibfnamefont
  {D.}~\bibnamefont {Schmitt}},\ }\href
  {http://dx.doi.org/10.1103/PhysRevE.83.066310} {\bibfield  {journal}
  {\bibinfo  {journal} {Phys. Rev. E}\ }\textbf {\bibinfo {volume} {83}},\
  \bibinfo {pages} {066310} (\bibinfo {year} {2011})}\BibitemShut {NoStop}%
\bibitem [{\citenamefont {Zimmerman}\ \emph {et~al.}(2014)\citenamefont
  {Zimmerman}, \citenamefont {Triana}, \citenamefont {Nataf},\ and\
  \citenamefont {Lathrop}}]{zimmerman14}%
  \BibitemOpen
  \bibfield  {author} {\bibinfo {author} {\bibfnamefont {D.~S.}\ \bibnamefont
  {Zimmerman}}, \bibinfo {author} {\bibfnamefont {S.~A.}\ \bibnamefont
  {Triana}}, \bibinfo {author} {\bibfnamefont {H.-C.}\ \bibnamefont {Nataf}}, \
  and\ \bibinfo {author} {\bibfnamefont {D.~P.}\ \bibnamefont {Lathrop}},\
  }\href {\doibase 10.1002/2013JB010733} {\bibfield  {journal} {\bibinfo
  {journal} {J. Geophys. Res.: Solid Earth}\ }\textbf {\bibinfo {volume}
  {119}},\ \bibinfo {pages} {4538} (\bibinfo {year} {2014})}\BibitemShut
  {NoStop}%
\bibitem [{\citenamefont {O'Connell}\ \emph {et~al.}(2001)\citenamefont
  {O'Connell}, \citenamefont {Kendrick}, \citenamefont {Nornberg},
  \citenamefont {Spence}, \citenamefont {Bayliss},\ and\ \citenamefont
  {Forest}}]{oconnell01}%
  \BibitemOpen
  \bibfield  {author} {\bibinfo {author} {\bibfnamefont {R.}~\bibnamefont
  {O'Connell}}, \bibinfo {author} {\bibfnamefont {R.}~\bibnamefont {Kendrick}},
  \bibinfo {author} {\bibfnamefont {M.}~\bibnamefont {Nornberg}}, \bibinfo
  {author} {\bibfnamefont {E.}~\bibnamefont {Spence}}, \bibinfo {author}
  {\bibfnamefont {A.}~\bibnamefont {Bayliss}}, \ and\ \bibinfo {author}
  {\bibfnamefont {C.}~\bibnamefont {Forest}},\ }in\ \href {\doibase
  10.1007/978-94-010-0788-7_7} {\emph {\bibinfo {booktitle} {Dynamo and
  Dynamics, a Mathematical Challenge}}}\ (\bibinfo  {publisher} {Springer},\
  \bibinfo {year} {2001})\ pp.\ \bibinfo {pages} {59--66}\BibitemShut {NoStop}%
\bibitem [{\citenamefont {Lathrop}\ \emph {et~al.}(2001)\citenamefont
  {Lathrop}, \citenamefont {Shew},\ and\ \citenamefont {Sisan}}]{lathrop01}%
  \BibitemOpen
  \bibfield  {author} {\bibinfo {author} {\bibfnamefont {D.~P.}\ \bibnamefont
  {Lathrop}}, \bibinfo {author} {\bibfnamefont {W.~L.}\ \bibnamefont {Shew}}, \
  and\ \bibinfo {author} {\bibfnamefont {D.~R.}\ \bibnamefont {Sisan}},\ }\href
  {\doibase 10.1088/0741-3335/43/12A/311} {\bibfield  {journal} {\bibinfo
  {journal} {Plasma Physics and controlled fusion}\ }\textbf {\bibinfo {volume}
  {43}},\ \bibinfo {pages} {A151} (\bibinfo {year} {2001})}\BibitemShut
  {NoStop}%
\bibitem [{\citenamefont {Mari{\'e}}\ \emph {et~al.}(2002)\citenamefont
  {Mari{\'e}}, \citenamefont {Petrelis}, \citenamefont {Bourgoin},
  \citenamefont {Burguete}, \citenamefont {Chiffaudel}, \citenamefont
  {Daviaud}, \citenamefont {Fauve}, \citenamefont {Odier},\ and\ \citenamefont
  {Pinton}}]{marie02b}%
  \BibitemOpen
  \bibfield  {author} {\bibinfo {author} {\bibfnamefont {L.}~\bibnamefont
  {Mari{\'e}}}, \bibinfo {author} {\bibfnamefont {F.}~\bibnamefont {Petrelis}},
  \bibinfo {author} {\bibfnamefont {M.}~\bibnamefont {Bourgoin}}, \bibinfo
  {author} {\bibfnamefont {J.}~\bibnamefont {Burguete}}, \bibinfo {author}
  {\bibfnamefont {A.}~\bibnamefont {Chiffaudel}}, \bibinfo {author}
  {\bibfnamefont {F.}~\bibnamefont {Daviaud}}, \bibinfo {author} {\bibfnamefont
  {S.}~\bibnamefont {Fauve}}, \bibinfo {author} {\bibfnamefont
  {P.}~\bibnamefont {Odier}}, \ and\ \bibinfo {author} {\bibfnamefont
  {J.}~\bibnamefont {Pinton}},\ }\href {http://hdl.handle.net/10171/1610}
  {\bibfield  {journal} {\bibinfo  {journal} {{Magnetohydrodynamics}}\ }\textbf
  {\bibinfo {volume} {38}},\ \bibinfo {pages} {163} (\bibinfo {year}
  {2002})}\BibitemShut {NoStop}%
\bibitem [{\citenamefont {{Monchaux}}\ \emph {et~al.}(2007)\citenamefont
  {{Monchaux}}, \citenamefont {{Berhanu}}, \citenamefont {{Bourgoin}},
  \citenamefont {{Moulin}}, \citenamefont {{Odier}}, \citenamefont {{Pinton}},
  \citenamefont {{Volk}}, \citenamefont {{Fauve}}, \citenamefont {{Mordant}},
  \citenamefont {{P\' etr\' elis}}, \citenamefont {{Chiffaudel}}, \citenamefont
  {{Daviaud}}, \citenamefont {{Dubrulle}}, \citenamefont {{Gasquet}},
  \citenamefont {{Mari\' e}},\ and\ \citenamefont {{Ravelet}}}]{monchaux07}%
  \BibitemOpen
  \bibfield  {author} {\bibinfo {author} {\bibfnamefont {R.}~\bibnamefont
  {{Monchaux}}}, \bibinfo {author} {\bibfnamefont {M.}~\bibnamefont
  {{Berhanu}}}, \bibinfo {author} {\bibfnamefont {M.}~\bibnamefont
  {{Bourgoin}}}, \bibinfo {author} {\bibfnamefont {M.}~\bibnamefont
  {{Moulin}}}, \bibinfo {author} {\bibfnamefont {P.}~\bibnamefont {{Odier}}},
  \bibinfo {author} {\bibfnamefont {J.-F.}\ \bibnamefont {{Pinton}}}, \bibinfo
  {author} {\bibfnamefont {R.}~\bibnamefont {{Volk}}}, \bibinfo {author}
  {\bibfnamefont {S.}~\bibnamefont {{Fauve}}}, \bibinfo {author} {\bibfnamefont
  {N.}~\bibnamefont {{Mordant}}}, \bibinfo {author} {\bibfnamefont
  {F.}~\bibnamefont {{P\' etr\' elis}}}, \bibinfo {author} {\bibfnamefont
  {A.}~\bibnamefont {{Chiffaudel}}}, \bibinfo {author} {\bibfnamefont
  {F.}~\bibnamefont {{Daviaud}}}, \bibinfo {author} {\bibfnamefont
  {B.}~\bibnamefont {{Dubrulle}}}, \bibinfo {author} {\bibfnamefont
  {C.}~\bibnamefont {{Gasquet}}}, \bibinfo {author} {\bibfnamefont
  {L.}~\bibnamefont {{Mari\' e}}}, \ and\ \bibinfo {author} {\bibfnamefont
  {F.}~\bibnamefont {{Ravelet}}},\ }\href {\doibase
  10.1103/PhysRevLett.98.044502} {\bibfield  {journal} {\bibinfo  {journal}
  {Phys. Rev. Lett.}\ }\textbf {\bibinfo {volume} {98}},\ \bibinfo {pages}
  {044502} (\bibinfo {year} {2007})}\BibitemShut {NoStop}%
\bibitem [{\citenamefont {{Berhanu}}\ \emph {et~al.}(2007)\citenamefont
  {{Berhanu}}, \citenamefont {{Monchaux}}, \citenamefont {{Fauve}},
  \citenamefont {{Mordant}}, \citenamefont {{P\' etr\' elis}}, \citenamefont
  {{Chiffaudel}}, \citenamefont {{Daviaud}}, \citenamefont {{Dubrulle}},
  \citenamefont {{Mari\' e}}, \citenamefont {{Ravelet}}, \citenamefont
  {{Bourgoin}}, \citenamefont {{Odier}}, \citenamefont {{Pinton}},\ and\
  \citenamefont {{Volk}}}]{berhanu07}%
  \BibitemOpen
  \bibfield  {author} {\bibinfo {author} {\bibfnamefont {M.}~\bibnamefont
  {{Berhanu}}}, \bibinfo {author} {\bibfnamefont {R.}~\bibnamefont
  {{Monchaux}}}, \bibinfo {author} {\bibfnamefont {S.}~\bibnamefont {{Fauve}}},
  \bibinfo {author} {\bibfnamefont {N.}~\bibnamefont {{Mordant}}}, \bibinfo
  {author} {\bibfnamefont {F.}~\bibnamefont {{P\' etr\' elis}}}, \bibinfo
  {author} {\bibfnamefont {A.}~\bibnamefont {{Chiffaudel}}}, \bibinfo {author}
  {\bibfnamefont {F.}~\bibnamefont {{Daviaud}}}, \bibinfo {author}
  {\bibfnamefont {B.}~\bibnamefont {{Dubrulle}}}, \bibinfo {author}
  {\bibfnamefont {L.}~\bibnamefont {{Mari\' e}}}, \bibinfo {author}
  {\bibfnamefont {F.}~\bibnamefont {{Ravelet}}}, \bibinfo {author}
  {\bibfnamefont {M.}~\bibnamefont {{Bourgoin}}}, \bibinfo {author}
  {\bibfnamefont {P.}~\bibnamefont {{Odier}}}, \bibinfo {author} {\bibfnamefont
  {J.-F.}\ \bibnamefont {{Pinton}}}, \ and\ \bibinfo {author} {\bibfnamefont
  {R.}~\bibnamefont {{Volk}}},\ }\href {\doibase 10.1209/0295-5075/77/59001}
  {\bibfield  {journal} {\bibinfo  {journal} {Europhys. Lett.}\ }\textbf
  {\bibinfo {volume} {77}},\ \bibinfo {pages} {59001} (\bibinfo {year}
  {2007})}\BibitemShut {NoStop}%
\bibitem [{\citenamefont {Spence}\ \emph {et~al.}(2006)\citenamefont {Spence},
  \citenamefont {Nornberg}, \citenamefont {Jacobson}, \citenamefont
  {Kendrick},\ and\ \citenamefont {Forest}}]{spence06}%
  \BibitemOpen
  \bibfield  {author} {\bibinfo {author} {\bibfnamefont {E.}~\bibnamefont
  {Spence}}, \bibinfo {author} {\bibfnamefont {M.}~\bibnamefont {Nornberg}},
  \bibinfo {author} {\bibfnamefont {C.}~\bibnamefont {Jacobson}}, \bibinfo
  {author} {\bibfnamefont {R.}~\bibnamefont {Kendrick}}, \ and\ \bibinfo
  {author} {\bibfnamefont {C.}~\bibnamefont {Forest}},\ }\href {\doibase
  10.1103/PhysRevLett.96.055002} {\bibfield  {journal} {\bibinfo  {journal}
  {{Phys. Rev. Lett.}}\ }\textbf {\bibinfo {volume} {{96}}},\ \bibinfo {pages}
  {055002} (\bibinfo {year} {{2006}})}\BibitemShut {NoStop}%
\bibitem [{\citenamefont {Nataf}(2013)}]{nataf13}%
  \BibitemOpen
  \bibfield  {author} {\bibinfo {author} {\bibfnamefont {H.-C.}\ \bibnamefont
  {Nataf}},\ }\href {http://dx.doi.org/10.1016/j.crhy.2012.12.002} {\bibfield
  {journal} {\bibinfo  {journal} {Comptes Rendus Physique}\ }\textbf {\bibinfo
  {volume} {14}},\ \bibinfo {pages} {248} (\bibinfo {year} {2013})}\BibitemShut
  {NoStop}%
\bibitem [{\citenamefont {Figueroa}\ \emph {et~al.}(2013)\citenamefont
  {Figueroa}, \citenamefont {Schaeffer}, \citenamefont {Nataf},\ and\
  \citenamefont {Schmitt}}]{figueroa13}%
  \BibitemOpen
  \bibfield  {author} {\bibinfo {author} {\bibfnamefont {A.}~\bibnamefont
  {Figueroa}}, \bibinfo {author} {\bibfnamefont {N.}~\bibnamefont {Schaeffer}},
  \bibinfo {author} {\bibfnamefont {H.-C.}\ \bibnamefont {Nataf}}, \ and\
  \bibinfo {author} {\bibfnamefont {D.}~\bibnamefont {Schmitt}},\ }\href
  {\doibase 10.1017/jfm.2012.551} {\bibfield  {journal} {\bibinfo  {journal}
  {J. Fluid Mech.}\ }\textbf {\bibinfo {volume} {716}},\ \bibinfo {pages} {445}
  (\bibinfo {year} {2013})}\BibitemShut {NoStop}%
\bibitem [{\citenamefont {{Nataf}}\ \emph {et~al.}(2006)\citenamefont
  {{Nataf}}, \citenamefont {{Alboussi\`ere}}, \citenamefont {{Brito}},
  \citenamefont {{Cardin}}, \citenamefont {{Gagni\`ere}}, \citenamefont
  {{Jault}}, \citenamefont {{Masson}},\ and\ \citenamefont
  {{Schmitt}}}]{nataf06}%
  \BibitemOpen
  \bibfield  {author} {\bibinfo {author} {\bibfnamefont {H.-C.}\ \bibnamefont
  {{Nataf}}}, \bibinfo {author} {\bibfnamefont {T.}~\bibnamefont
  {{Alboussi\`ere}}}, \bibinfo {author} {\bibfnamefont {D.}~\bibnamefont
  {{Brito}}}, \bibinfo {author} {\bibfnamefont {P.}~\bibnamefont {{Cardin}}},
  \bibinfo {author} {\bibfnamefont {N.}~\bibnamefont {{Gagni\`ere}}}, \bibinfo
  {author} {\bibfnamefont {D.}~\bibnamefont {{Jault}}}, \bibinfo {author}
  {\bibfnamefont {J.-P.}\ \bibnamefont {{Masson}}}, \ and\ \bibinfo {author}
  {\bibfnamefont {D.}~\bibnamefont {{Schmitt}}},\ }\href {\doibase
  10.1080/03091920600718426} {\bibfield  {journal} {\bibinfo  {journal}
  {Geophys. Astrophys. Fluid Dyn.}\ }\textbf {\bibinfo {volume} {100}},\
  \bibinfo {pages} {281} (\bibinfo {year} {2006})}\BibitemShut {NoStop}%
\bibitem [{\citenamefont {Stefani}\ and\ \citenamefont
  {Gerbeth}(2000)}]{stefani00}%
  \BibitemOpen
  \bibfield  {author} {\bibinfo {author} {\bibfnamefont {F.}~\bibnamefont
  {Stefani}}\ and\ \bibinfo {author} {\bibfnamefont {G.}~\bibnamefont
  {Gerbeth}},\ }\href {\doibase 10.1088/0266-5611/16/1/301} {\bibfield
  {journal} {\bibinfo  {journal} {Inverse Problems}\ }\textbf {\bibinfo
  {volume} {16}},\ \bibinfo {pages} {1} (\bibinfo {year} {2000})}\BibitemShut
  {NoStop}%
\bibitem [{\citenamefont {Schaeffer}(2013)}]{schaeffer13}%
  \BibitemOpen
  \bibfield  {author} {\bibinfo {author} {\bibfnamefont {N.}~\bibnamefont
  {Schaeffer}},\ }\href {\doibase 10.1002/ggge.20071} {\bibfield  {journal}
  {\bibinfo  {journal} {Geochem. Geophys. Geosys.}\ }\textbf {\bibinfo {volume}
  {14}},\ \bibinfo {pages} {751} (\bibinfo {year} {2013})}\BibitemShut
  {NoStop}%
\bibitem [{\citenamefont {Tarantola}\ and\ \citenamefont
  {Valette}(1982)}]{tarantola82}%
  \BibitemOpen
  \bibfield  {author} {\bibinfo {author} {\bibfnamefont {A.}~\bibnamefont
  {Tarantola}}\ and\ \bibinfo {author} {\bibfnamefont {B.}~\bibnamefont
  {Valette}},\ }\href {http://dx.doi.org/10.1029/RG020i002p00219} {\bibfield
  {journal} {\bibinfo  {journal} {Reviews of Geophysics}\ }\textbf {\bibinfo
  {volume} {20}},\ \bibinfo {pages} {219} (\bibinfo {year} {1982})}\BibitemShut
  {NoStop}%
\bibitem [{\citenamefont {Ferraro}(1937)}]{ferraro37}%
  \BibitemOpen
  \bibfield  {author} {\bibinfo {author} {\bibfnamefont {V.}~\bibnamefont
  {Ferraro}},\ }\href {http://dx.doi.org/10.1093/mnras/97.6.458} {\bibfield
  {journal} {\bibinfo  {journal} {Mon. Not. Roy. Astron. Soc.}\ }\textbf
  {\bibinfo {volume} {97}},\ \bibinfo {pages} {458} (\bibinfo {year}
  {1937})}\BibitemShut {NoStop}%
\bibitem [{\citenamefont {Wicht}(2014)}]{wicht14}%
  \BibitemOpen
  \bibfield  {author} {\bibinfo {author} {\bibfnamefont {J.}~\bibnamefont
  {Wicht}},\ }\href {\doibase 10.1017/jfm.2013.545} {\bibfield  {journal}
  {\bibinfo  {journal} {J. Fluid Mech.}\ }\textbf {\bibinfo {volume} {738}},\
  \bibinfo {pages} {184} (\bibinfo {year} {2014})}\BibitemShut {NoStop}%
\bibitem [{\citenamefont {Cabanes}\ \emph {et~al.}(2014)\citenamefont
  {Cabanes}, \citenamefont {Schaeffer},\ and\ \citenamefont
  {Nataf}}]{Cabanes14b}%
  \BibitemOpen
  \bibfield  {author} {\bibinfo {author} {\bibfnamefont {S.}~\bibnamefont
  {Cabanes}}, \bibinfo {author} {\bibfnamefont {N.}~\bibnamefont {Schaeffer}},
  \ and\ \bibinfo {author} {\bibfnamefont {H.-C.}\ \bibnamefont {Nataf}},\
  }\href {\doibase 10.1103/PhysRevLett.113.184501} {\bibfield  {journal}
  {\bibinfo  {journal} {Phys. Rev. Lett.}\ }\textbf {\bibinfo {volume} {113}},\
  \bibinfo {pages} {184501} (\bibinfo {year} {2014})}\BibitemShut {NoStop}%
\bibitem [{\citenamefont {Candes}\ and\ \citenamefont {Romberg}()}]{candes05}%
  \BibitemOpen
  \bibfield  {author} {\bibinfo {author} {\bibfnamefont {E.}~\bibnamefont
  {Candes}}\ and\ \bibinfo {author} {\bibfnamefont {J.}~\bibnamefont
  {Romberg}},\ }\href {http://users.ece.gatech.edu/~justin/l1magic/} {\bibinfo
  {journal} {http://users.ece.gatech.edu/~justin/l1magic/}\ }\BibitemShut
  {NoStop}%
\end{thebibliography}%

\end{document}